\documentclass[fleqn,usenatbib]{mnras}

\usepackage{newtxtext,newtxmath}

\usepackage[T1]{fontenc}
\usepackage{ae,aecompl}

\usepackage{amsmath}
\usepackage{graphicx}
\usepackage{natbib}
\usepackage{color}
\usepackage{soul}

\newcommand{\lin}{LIN 358}
\newcommand{\ha}{H$\alpha$}
\newcommand{\hb}{H$\beta$}
\newcommand{\fe}{[Fe \textsc{x}] 6374 \AA}

\title[The symbiotic binary LIN 358]{LIN 358: A symbiotic binary accreting above the steady hydrogen fusion limit}

\author[J. Kuuttila et al.]{
J. Kuuttila$^{1}$,
M. Gilfanov$^{1,2}$,
T. E. Woods$^{3}$,
I. R. Seitenzahl$^{4}$,
and A. J. Ruiter$^{4}$
\\
$^{1}$Max Planck Institute for Astrophysics, Karl-Schwarzschild-Str. 1, Garching b. M\"unchen 85741, Germany\\
$^{2}$Space Research Institute, Profsoyuznaya 84/32, 117997, Moscow, Russia\\
$^{3}$National Research Council of Canada, Herzberg Astronomy \& Astrophysics Research Centre, \\  5071 West Saanich Road, Victoria, BC V9E 2E7, Canada\\
$^{4}$School of Science, University of New South Wales, Australian Defence Force Academy, Canberra, ACT 2600, Australia\\
}

\date{Accepted XXX. Received YYY; in original form ZZZ}

\pubyear{2020}

\begin{document}
\label{firstpage}
\pagerange{\pageref{firstpage}--\pageref{lastpage}}
\maketitle

\begin{abstract}
Symbiotic binaries are long period interacting binaries consisting of a white dwarf (WD) accreting material from a cool evolved giant star via stellar winds. 
In this paper we study the symbiotic binary LIN 358 located in the SMC. We have observed LIN 358 with the integral field spectrograph WiFeS and obtained its line emission spectrum. 
With the help of the plasma simulation and spectral synthesis code \textsc{Cloudy}, we have constructed a 2D photo-ionisation model of LIN 358.
From comparison with the observations, we have determined the colour temperature of the WD in LIN 358 to be 19 eV, its bolometric luminosity $L = (1.02 \pm 0.15) \times 10^{38}$ erg s$^{-1}$, and the mass-loss rate from the donor star to be $ 1.2 \times 10^{-6}$ M$_{\odot}$ yr$^{-1}$. 
Assuming a solar H to He ratio in the wind material, a lower limit to the accreted mass fraction in LIN 358 is 0.31.
The high mass-accretion efficiency of a wind Roche lobe overflow implies that the WD is accreting above the upper boundary of stable hydrogen fusion and thus growing in mass with the maximal rate of $\approx 4 \times 10^{-7}$ M$_{\odot}$ yr$^{-1}$. This causes the WD photosphere to expand, which explains its low colour temperature. 
Our calculations show that the circumstellar material in LIN 358 is nearly completely ionized except for a narrow cone around the donor star, and that the WD emission is freely escaping the system. However, due to its low colour temperature, this emission can be easily attenuated by even moderate amounts of neutral ISM. We speculate that other symbiotic systems may be operating in a similar regime, thus explaining the paucity of observed systems.
\end{abstract}




\section{Introduction}

Symbiotic binaries, or symbiotic stars, are interacting binaries consisting of a hot compact object, usually a white dwarf (WD), though neutron stars are also possible, accreting material from a cool evolved giant. The donor star can be either a normal red giant (S-type) or a Mira type variable embedded in an optically thick dust shell (D-type). Symbiotic binaries have the longest orbital separations and periods among the interacting binaries, with the periods ranging up to tens of years in D-type binaries \citep[e.g. 43.6 years for R Aqr;][]{Mikolajewska09}. 
To date, there are 257 confirmed symbiotic binaries in the Milky Way and 66 extra-galactic objects \citep{Akras19}, which is much less than predicted by theoretical estimates, which range from $10^3$ \citep[e.g.][]{Lu06, Yungelson10} to a few times $10^5$ \citep{Magrini03}.

Symbiotic binaries are surrounded by a complex circumstellar environment, the result of the hot ionizing compact object being embedded in the dense neutral wind of the giant star, with both ionized and neutral regions present as well as dust forming regions, accretion discs and possibly jets. The vast array of differing conditions makes symbiotic binaries excellent test cases of late stages of stellar evolution and binary interactions. In addition,  symbiotic binaries have been discussed in the context of the problem of  progenitors of Type Ia supernovae (SNe Ia)  \citep[see][for a review]{Maoz14}, both in single and double degenerate channels, and also as a possible channel of neutron star formation via accretion-induced collapse \citep[AIC;][]{Nomoto91, Wang18}.  For a thorough review of symbiotic binaries, see \citet{Mikolajewska12}. 

Symbiotic binaries may also exhibit other phenomena usually associated with accreting white dwarfs, such as thermonuclear novae, either slow or recurrent, where the hydrogen accreted from the giant onto the WD burns into heavier elements, either in a long outburst or in a short flash. These novae typically cause some amount of matter to be ejected from the WD, which may prevent the WD from growing in mass \citep{Wolf13}, except in the case of short recurrence time novae, where the WD can retain a significant amount of the accreted material due to the higher interior temperature and thus less explosive burning \citep{Hillman16}. 
If the WD in the binary is accreting material steadily, the system may appear as a super-soft X-ray binary (SSS), which is characterised by effective temperatures of $10^{5-6}$ K and luminosities of $10^{37-38}$ erg s$^{-1}$ \citep{Greiner00}. In the SSS phase the accreted hydrogen is burned steadily on the surface of the white dwarf, which allows the mass of the WD to grow efficiently \citep{vandenHeuvel92}. 

LIN 358, also known as RX J0059.1-7505, is an S-type symbiotic binary consisting of a WD and an asymptotic giant branch (AGB) star and located in the outskirts of the Small Magellanic Cloud (SMC) at coordinates RA = 00h 59m 12.3s, Dec = -75$^{\circ}$ 05$'$ 17.6$''$. It was first discovered by \citet{Lindsay61} and characterised as a symbiotic binary by \citet{Walker83} using optical observations. \citet{Muerset97} analysed the ROSAT PSPC observations of LIN 358 and classified it as a super-soft X-ray source based on its X-ray spectrum. \citet{Kahabka06} observed LIN 358 with XMM-Newton and obtained an effective temperature of $T_h = 227.5 \pm 30$ kK and luminosity $L_h = 1.0 \times 10^{38}$ erg s$^{-1}$ for the hot component of the binary from their black-body fit to the super-soft component (0.13 -- 1.0 keV). LIN 358 was also studied by \citet{Skopal15a}, who used multi-wavelength modelling of the spectral energy distribution to determine the effective temperature $T_h = 250 \pm 10$ kK and luminosity $\mathrm{log_{10}(}L_h\mathrm{)} = 38.03 \pm 0.11$ erg s$^{-1}$ for the WD, in agreement with the previous X-ray analysis. 
In addition, \citet{Skopal15a} derived the effective temperature $T_{\mathrm{g}} = 4000 \pm 200$ K, bolometric luminosity $L_g = (2.8 \pm 0.8) \times 10^{37} \, (d/\mathrm{60 kpc})^2$ erg s$^{-1}$, and the radius $R_g = 178 \, (d/\mathrm{60 kpc})^2$ $\mathrm{R_{\odot}}$ for the giant star. \citet{Skopal15a} determined the properties of the giant by matching the photometric \textit{BVJHK} flux points with a synthetic spectrum calculated for $T_{\mathrm{g}} = 4000 \pm 200$ K from a grid of giant atmosphere models calculated by \citet{Hauschildt99}. This is possible due to the fact that in the near-IR the emission from the AGB star dominates over the emission from both the WD, which peaks in the UV, and the ionized nebula, which is most prominent in the optical lines \citep{Skopal05}. Using these parameters \citet{Skopal15a} classified the giant star in LIN 358 to be a type K5 Ib supergiant. 

Apart from the temperatures and luminosities, not much is known about this system. In particular, the mass accretion rate and the mass of the WD, which are the most important parameters in determining the much speculated viability of symbiotic binaries as Type Ia supernova progenitors, are still unknown. In this work we determine the temperature, mass, and mass accretion rate of LIN 358 by using our new optical spectroscopic observations together with photoionization calculations performed with the spectral synthesis code \textsc{Cloudy}. 
This paper is organised as follows: In Sec.~\ref{sec:ov} we review the overall properties and geometry of LIN 358. In Sec.~\ref{sec:obs} we describe our observations and the data reduction procedure, in Sec.~\ref{sec:data} we present the data, and in Sec.~\ref{sec:sim} we describe the \textsc{Cloudy} simulations. In Sec.~\ref{sec:results} we present our results and discuss the possible implications.

\section{LIN 358 overview}\label{sec:ov}

\subsection{Mass estimation}\label{sec:mass}

\citet{Skopal15a} classified the cool giant in \lin\ to be a K5 Ib type supergiant by matching the photometric BVJHK flux points of \citet{Muerset1996} to the spectral models calculated by \citet{Hauschildt99}. These models were calculated for a mass of 5 M$_{\odot}$, which is a quite typical mass for a cool supergiant AGB star of this spectral type \citep{Hohle10}, so we have adopted 5 M$_{\odot}$ to be the mass of the giant star in \lin .

The mass of the giant star is important in estimating a lower limit for the WD mass, assuming that both of the stars in the binary were born at the same time. If the current giant has a mass of 5 M$_{\odot}$, then the progenitor star of the WD should have had larger initial mass in order to evolve before the current giant star. According to the initial mass -- WD mass relationship of \citet{Cummings18}, a star with an initial mass \textgreater 5 M$_{\odot}$ should create a WD with mass \textgreater 1 M$_{\odot}$. In addition, the X-ray spectral fits of LIN358 by \citet{Orio07} indicate a WD mass $> 1.18 $ M$_{\odot}$. In the rest of the paper we have assumed a mass of 1 M$_{\odot}$ for the WD. We note, however, that our results are only weakly sensitive to the adopted masses, because they affect only the orbital separation of the binary (Sec.~\ref{sec:orb}), which in turn does not affect our results significantly (see Sec.~\ref{sec:discussion}).

The mass of the WD has important implications on the composition of the WD, because the maximum mass of a newborn carbon and oxygen (CO) rich WDs is $\sim 1.2$ M$_{\odot}$. AGB stars with masses $\gtrsim 6 $ M$_{\odot}$ produce higher mass WDs, which are believed to be formed as oxygen, neon, and magnesium (ONeMg) rich. This in turn affects the possible end results of the binary, because only CO WDs are thought to produce SNe Ia and ONeMg WDs are believed to form neutron stars via AIC \citep{Nomoto84, Nomoto91}. 

The mass estimates can be affected also by the pre-WD binary evolution. When the current WD went through the AGB phase, it lost the majority of its mass through stellar winds, and  fraction of the mass lost may have been accreted by the second star, thus skewing the current mass ratio \citep{vandenHeuvel94}. A careful modelling of the binary properties is required to examine this problem and is thus outside the scope of this paper. This effect, however, is likely not significant for our results, because the typical mass accreted mass is only $\sim 10$ \% of the mass lost by the former AGB star. In the rest of the paper we have assumed that there were not any significant interactions before the current evolutionary stage.

\subsection{Orbital parameters}\label{sec:orb}

Despite the large orbital periods, the WDs in symbiotic binaries are often accreting material efficiently from the AGB star. 
However, in most symbiotic binaries, the orbital separation is too large for the standard Roche lobe overflow (RLOF) scenario. In addition, interaction via RLOF from an AGB star with a deep convective envelope may often lead to unstable mass transfer \citep[][though see e.g., \citealt{WI11}]{Hjellming87, Chen08} and a common envelope phase \citep{Paczynski76}.

AGB stars are known to have strong stellar winds with mass-loss rates on the order of $10^{-8} - 10^{-5}$ M$_{\odot}$ yr$^{-1}$ and low velocities on the order of $5 - 30$ km s$^{-1}$ \citep{Hoefner18}. Therefore, instead of RLOF, the WD is assumed to accrete material from the wind of the donor star. 
However, the standard Bondi-Hoyle-Lyttleton \citep[BHL;][]{Hoyle39, Bondi44} wind accretion scenario often fails to explain the required mass accretion rates. 
The BHL description is a good approximation of the wind accretion when the outflow velocity is fast compared to the orbital velocity, which is not the case for typical AGB winds. 

Recent simulations suggest a new mode of mass transfer, called wind Roche lobe overflow \citep[WRLOF;][]{Mohamed07, Mohamed12}, where the star itself does not fill the Roche lobe, but the stellar wind is confined in the Roche lobe, because the wind acceleration radius $R_{d}$ is larger than the Roche lobe radius $R_{L,1}$. In this situation the wind is focused towards the orbital plane \citep{deValBorro09}, allowing an efficient mass transfer through the Lagrangian 1 (L1) point. In WRLOF the mass-transfer rate may exceed the estimated rates from the simple BHL accretion by up to 2 orders of magnitude. 

The conditions necessary for WRLOF can be estimated with the ratio $R_{d} / R_{L,1}$. The Roche-lobe radius of the donor star $R_{L,1}$ depends on the mass ratio $q$ and the binary separation $a$, and can be estimated as \citep{Eggleton83}:
\begin{equation}\label{eq:roche}
    R_{L,1} \,\, = \,\, a \, \times \, \frac{0.49 q^{2/3}}{0.6 q^{2/3} + \mathrm{ln}(1 + q^{1/3})}. 
\end{equation}
In AGB stars, the stellar winds are driven by dust \citep{Hoefner15, Hoefner18}, which means that the wind acceleration radius coincides with the dust condensation radius, i.e. the radius where the gas is cooled enough to form dust grains. 
This radius can be estimated as \citep{Lamers99, Hoefner07}:
\begin{equation}\label{eq:rd}
     R_d \,\, = \,\, \frac{1}{2} R_* \left( \frac{T_{\mathrm{g}}}{T_{\mathrm{cond}}} \right) ^{\frac{4+p}{2}},
\end{equation}
where $R_*$ is the stellar radius and $T_g$ is the temperature of donor star. The condensation temperature $T_{\mathrm{cond}}$ and the exponent $p$ are characteristics of the grain material and depend on the chemical composition. 

\lin\ is an S-type symbiotic binary \citep{Muerset1996}, which means the infrared emission is dominated by the stellar continuum and not the dust emission as in D-type binaries. The dust grains are important in driving the stellar wind \citep{Hoefner18}, but the emission from the WD will later destroy the dust grains and ionize most of the wind (see Sec.~\ref{sec:csmabs}).

The atmosphere of the AGB star in LIN358 is O-rich \citep{Muerset1996}, meaning the condensing grains are mainly various silicates. The exact nature of the O-rich condensates is still debated, but following \citet{Bladh12, Bladh15, Hoefner16, Hoefner18}, for the most efficient silicate grains $p \approx -1.0$ and $T_{\mathrm{cond}} \approx 1100$ K. Using these values in Eq.~(\ref{eq:rd}), we get $R_d = 617.2 $ R$_{\odot}$ ($\approx 2.9$ AU). 

Next, we can estimate the orbital separation $a$ in \lin\ by using the ratio of the wind acceleration radius and the Roche lobe radius of the donor $R_{d} / R_{L,1}$. Based on the hydrodynamical simulations of \citet{Mohamed12, Abate13}, the maximal accretion efficiency is reached at $R_{d} / R_{L,1} = 1.5$. Using this with the Eq.~(\ref{eq:roche}), we can calculate the semi-major axis to be $a = 3.7$ AU, which corresponds to a orbital period of $\approx 2.9$ years.
This is well within the range of typical periods for S-type symbiotic binaries ($\sim$ 1 -- 6 years; \citealt{Gromadzki13}). 

In the rest of the paper we have used the $a = 3.7$ AU, but we note that our results are not particularly sensitive to this chosen value (see Sec.~\ref{sec:discussion}).

\begin{figure*}
\centering
\includegraphics[width=0.97\textwidth]{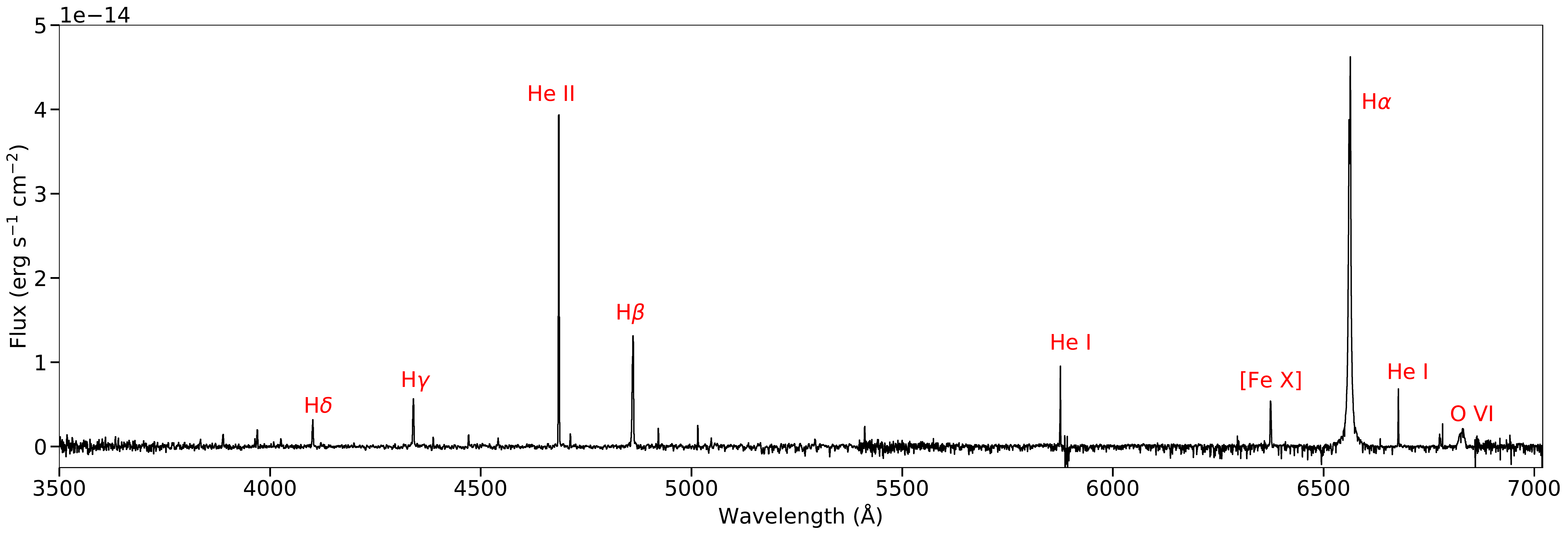}
\caption{The spectrum of \lin\ as observed with WiFeS. }
\label{fig:spectrum}
\end{figure*}


\section{Observations}\label{sec:obs}

We observed \lin\ on the nights of 2018 November 04--05 (P.I.: Seitenzahl; Proposal ID: 4180034) with the Wide Field Spectrograph (WiFeS) mounted on the Australian National University 2.3\,m telescope at the Siding Spring Observatory. We present only a short summary of the data reduction method, which is described in detail by \citet{Dopita16} and \citet{Ghavamian17}.

The WiFeS is a double-beam spectrograph which provides simultaneous and independent channels for both the blue (3500--5700 \AA ) and red (5300--7000 \AA ) wavelength ranges. We used the B3000 and R7000 gratings which means the spectral resolution in the blue wavelength range is R = 3000 ($\Delta v \approx$ 100 km s$^{-1}$) and in the red R = 7000 ($\Delta v \approx$ 45 km s$^{-1}$). 
The observations were performed in the `binned mode', which provided us a field of view of 25 $\times$ 35 spatial pixels (or spaxels), each of them $1'' \times 1''$ in angular size. This correspond to a field of view of 7.3 pc $\times$ 10.2 pc assuming a distance of 60 kpc to the SMC.\footnote{\citet{Scowcroft16} found the the distance to SMC to be $62.0 \pm 0.3$ kpc, but for easier comparison with previous results we have assumed a 60 kpc distance in this paper.}
LIN 358 was observed in one pointing which was offset from the source by $15''$ in the axis of the wider side of the WiFeS field of view (35 spaxels, or $35''$). The observations consisted of $2 \times 1800$s on source exposures and $2 \times 900$s blank sky exposures, which were scaled and subtracted from the two co-added frames. 

The data were reduced with the \textsc{pywifes} v0.7.3 pipeline \citep{Childress14ascl,Childress14}, which provided us a wavelength calibrated, sensitivity corrected, and photometrically calibrated data cube. 
The data were dereddened using the extinction curves for SMC bar of \citet{Weingartner01} with an assumed carbon abundance of zero and using the column density $N_H = 7.6 \times 10^{20}$ cm$^{-2}$ obtained by \citet{Kahabka06} from blackbody fits to the \textit{XMM-Newton} data. 
LIN 358 is located in the outskirts of the SMC, not in the bar, but given the lack of an available extinction curve for this region, we use the curve for the SMC bar. This does not affect the results significantly because the dereddening factor for e.g. the He \textsc{ii} 4686 \AA\ line is only $\approx 1.04 $, so our measured fluxes may be underestimated at most by a few per cent, which is of the same order as the measured flux errors. 

In addition, we corrected the data for the redshift of 185.2 km s$^{-1}$, which was measured from several narrow emission lines (SMC heliocentric radial velocity is 145.6 km s$^{-1}$; \citealt{McConnachie12}), and subtracted the continuum emission using the Locally Weighted Scatterplot Smoothing algorithm (LOWESS; \citealt{Cleveland79}) similarly to \citet{Vogt17a, Vogt17b}.


\section{Data}\label{sec:data}

\begin{figure*}
\centering
\includegraphics[width=0.95\textwidth]{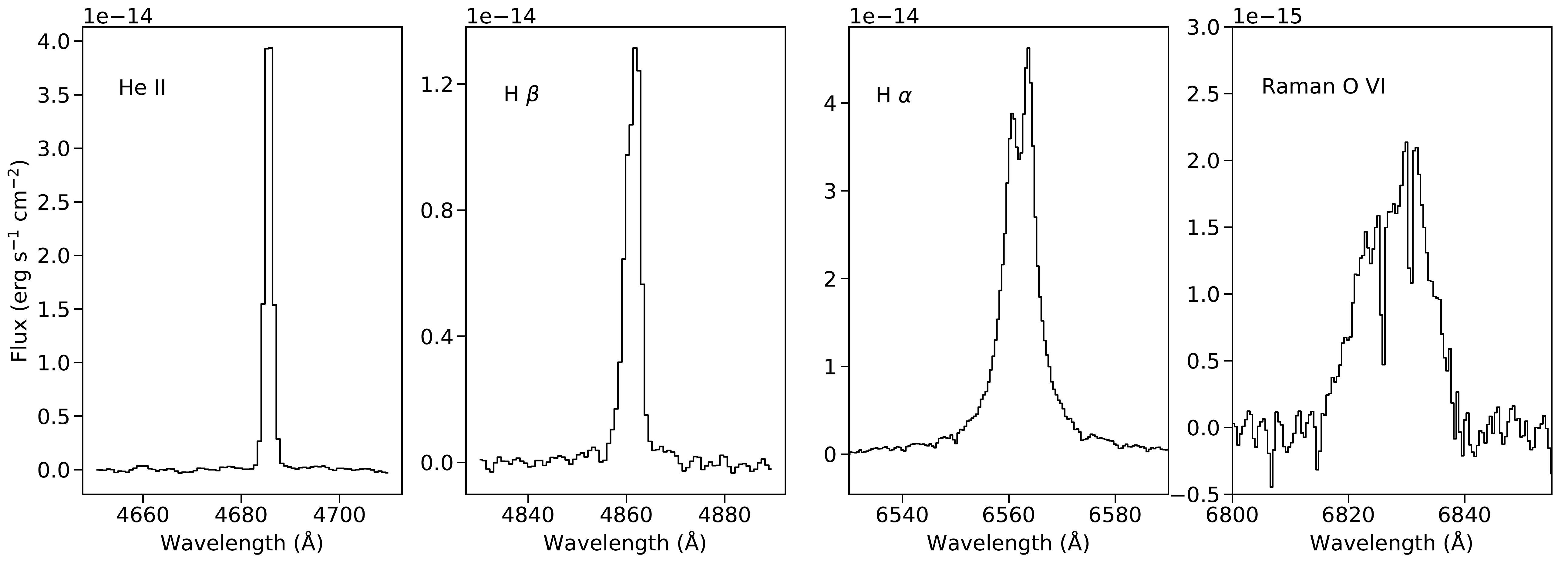}
\caption{The observed line profiles from left to right: He \textsc{ii} 4686\AA , \hb , \hb , and the Raman scattered O \textsc{vi} 6830 \AA .}
\label{fig:Ha_Hb}
\end{figure*}

After the data reduction process described in Sec.~\ref{sec:obs} we extracted the total source spectrum from the blue and red data cubes. The overall spectrum is show in Fig.~\ref{fig:spectrum}. The spectrum consists of various emission lines: H Balmer series from H$\alpha$ to the $n = 10 \rightarrow 2$ transition; various He \textsc{i} and He \textsc{ii} lines, most notably He \textsc{ii} 4686 \AA , He \textsc{i} 5876 \AA , and He \textsc{i} 6678 \AA ; \fe , and the [O \textsc{vi}] 6830 \AA\ Raman scattered feature. The usual nebular lines, such as [O \textsc{iii}] 4959, 5007 \AA\ and [S \textsc{ii}] 6716, 6731 \AA\ are notably absent in LIN 358. 
In our analysis we will focus mainly on the brightest observed lines, whose line luminosities are listed in Table~\ref{table:lums} and whose properties are outlined below.

\begin{table}
\caption{List of the luminosities of the brightest observed emission lines.}
\begin{tabular}{lc}
\multicolumn{1}{c}{Emission line}&\multicolumn{1}{c}{Luminosity $\times$ 10$^{33}$ (erg s$^{-1}$)}\\ \hline
He \textsc{ii} 4686 & 38.0 $\pm$ 1.2 \\
H $\beta$ & 23.1 $\pm$ 2.7  \\ 
He \textsc{i} 5876 & 6.2 $\pm$ 0.3  \\
$[$Fe \textsc{x}$]$ 6374 & 7.4 $\pm$ 0.4  \\
H $\alpha$ & 112.3 $\pm$ 8.1  \\ 
He \textsc{i} 6678 & 4.3 $\pm$ 0.2  \\ \hline
\end{tabular}
\label{table:lums}
\end{table}

\subsection{He II 4686 \AA}

The He \textsc{ii}  4686 \AA\ is the second brightest emission line in the spectrum with a luminosity L = $3.8 \pm 0.12 \times 10^{34}$ erg s$^{-1}$. This line comes from the n$ = 4 \rightarrow 3$ transition and is the brightest He \textsc{ii} line in the optical. The high ionization potential of He \textsc{ii} (54.4 eV) requires a hot ($\gtrsim$ 10$^{5}$ K) ionizing source, which makes this emission line a clear and important signature of an accreting white dwarf with steady nuclear burning \citep{Rappaport94}. For this reason this emission line has been used extensively in previous accreting WD and Type Ia supernova studies \citep{Woods13, Johansson14, Chen15, Woods16, Kuuttila19}. 
Other observed He \textsc{ii} lines are at 4199 \AA , 4541 \AA , and 5411 \AA .

\subsection{He I lines} 

The two brightest He \textsc{i} lines are at 5876 \AA\ and 6678 \AA , which are produced by the singlet transition 3$^1$D $\rightarrow$ 2$^1$P$^0$ and the triplet transition 3$^3$D $\rightarrow$ 2$^3$P$^0$, respectively. 
Other observed He \textsc{i} lines include the triplet lines at 3889 \AA\ and 4471 \AA , and the singlet lines at 3965 \AA , 4922 \AA , and 5016 \AA . 

Due to the high meta-stability of the He \textsc{i} excited level 2$^3$S, and to a lesser extent 2$^1$S, collisional effects play a significant role in the production of He \textsc{i} emission lines. Of the recombinations to excited levels of He \textsc{i}, approximately one fourth are to the singlet levels and three fourths are to the triplet levels, all of which eventually cascade down to the meta-stable 2$^3$S level through radiative transitions \citep{Osterbrock06}. The 2$^3$S level can decay to the ground state by emitting a photon, but at densities $\gtrsim 10^4$ cm$^{-3}$ most of the 2$^3$S states are depopulated by collisional transitions, for example to the singlet level 2$^1$S and triplet 2$^3$P$^0$ \citep{Bray00}. These collisional effects become even more important in the high densities and temperatures of symbiotic binaries. Calculating the luminosities of these lines thus requires a full treatment of all radiative and collisional processes.

\subsection{Balmer lines}

The \ha\ and \hb\ emission lines are among the most important astrophysical lines and typically are very well understood. However, in the case of \lin , the high density makes the treatment of these lines quite difficult. 
In lower density environments \citep[e.g., n $\ll 10^{6} $ cm$^{-3}$, see ][]{Hummer1987}, 
\ha\ and \hb\ emissivities can be treated using the simple Case B approximation \citep{Osterbrock06}. In the high density environment of symbiotic binaries, however, this is not the case, which is evident from the observed line ratio \ha /\hb\ $\approx$ 4.9, compared to the Case B line ratio \ha /\hb\ $\sim$ 3. The high densities and large optical depths in the nebulae around symbiotic binaries can cause the gas to become optically thick in the Balmer lines and self-absorption to occur.
This will drastically change the Balmer line ratios, as has been observed in some active galactic nuclei \citep{Netzer75} and also other symbiotic binaries \citep{Davidsen77}.
These conditions require a full radiative transfer treatment to fully model the line luminosities and ratios. 

The observed \ha\ and \hb\ line profiles are shown in Fig.~\ref{fig:Ha_Hb}. \ha\ requires four Gaussians components to be explained well: two narrow peaks, an intermediate component, and very broad ($\sim 2000$ km/s) wings. These kind of emission line profiles have been previously observed in many planetary nebulae and symbiotic binaries \citep[e.g.][]{Arrieta03, Chang18}.
There are several different possible formation mechanisms for the broad wings, e.g. optically thin stellar winds from the hot component \citep{Skopal06}, Thomson scattering by free electrons \citep{Sekeras12}, and Raman scattering of Ly$\beta$ photons \citep{Nussbaumer89, Lee00}. The Raman scattering is favoured by the fact that there are no clear broad components in the other hydrogen lines, because with Thomson wings \ha\ and \hb\ would have the same width that is proportional to $T_e ^{1/2}$, which is not the case in \lin. However, according to simulations of \citet{Chang18}, the Raman wings of \ha\ are about three times wider than the Raman wings of \hb\ due to the different cross sections for Ly$\beta$ and Ly$\gamma$, which fits the picture of \lin. In addition, the observed O \textsc{vi} 6830 \AA\ Raman feature shows that the conditions for Raman scattering are met, so it is reasonable to assume that the Balmer lines include a contribution from Raman scattering as well. 

\begin{figure}
\centering
\includegraphics[width=0.47\textwidth]{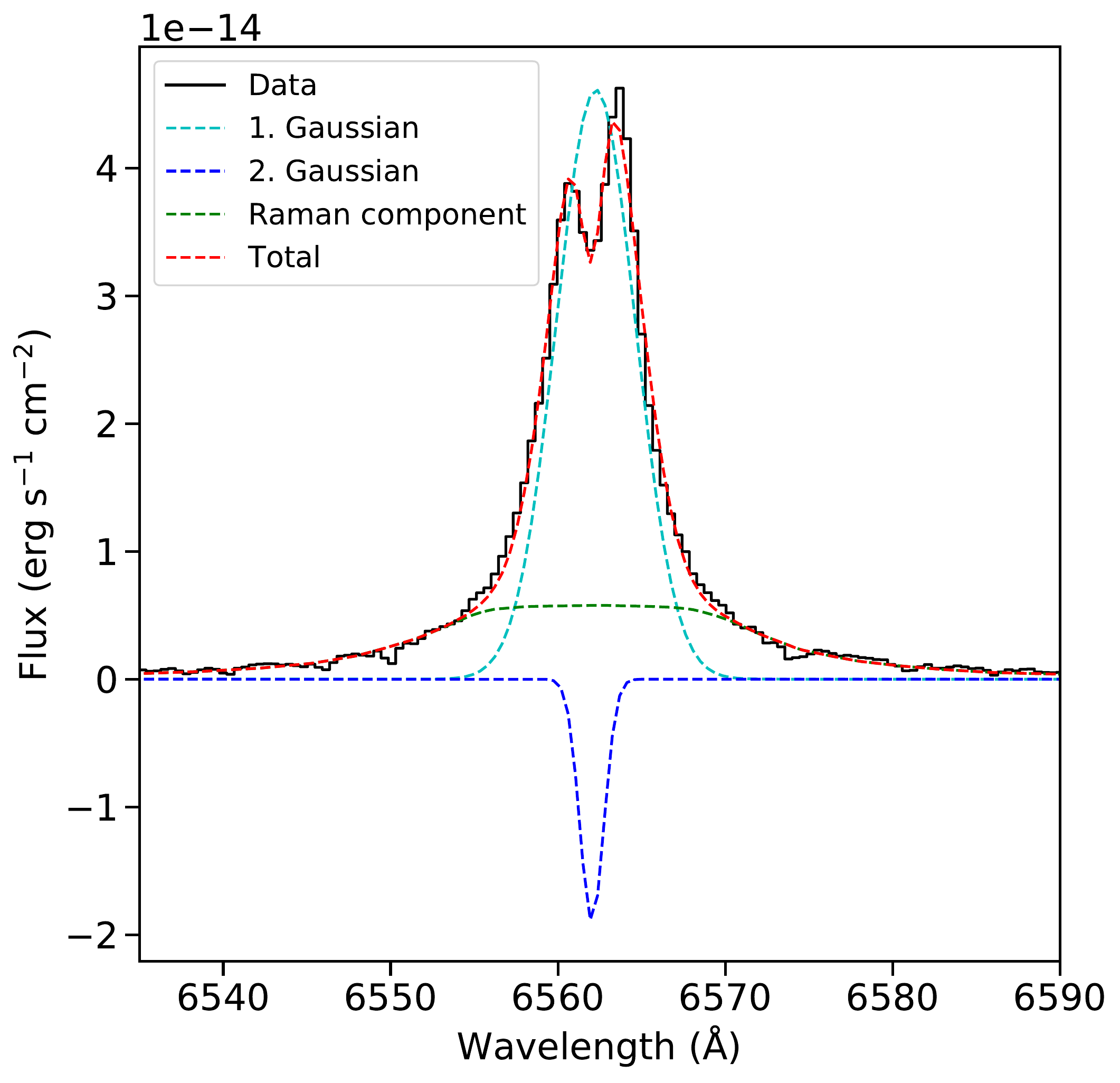}
\caption{The \ha\ line profile with fitted components. The black line is the observed data, green dashed line is the simulated line profile of \citet[][Fig. 6]{Chang18} for N$_{\mathrm{H\textsc{i}}}$ = 10$^{20}$ cm$^{-2}$, in cyan and blue dashed lines are the two Gaussian components, and in the red is the sum of all three. }
\label{fig:Ha_fit}
\end{figure}

In order to estimate the Balmer emission coming from the ionized region around the WD, we modelled the Raman scattered component with the simulated line profile of \citet[][Fig. 6]{Chang18} for N$_{\mathrm{H\textsc{i}}}$ = 10$^{20}$ cm$^{-2}$. After subtracting the Raman scattered component, the resulting line profile can be well fitted with two Gaussians, one emission line and one absorption line, see Fig.~\ref{fig:Ha_fit}.

\subsection{[Fe X] 6374 \AA}\label{sec:fex}

The coronal \fe\ line is the only forbidden line present in our data. This lines comes from the $^2$P$_{1/2} \rightarrow ^2$P$_{3/2}$ transition of Fe$^{9+}$. The ionisation energy of Fe$^{9+}$ is 233.6 eV, which makes this line very dependent on the temperature and thus a good, almost model independent test case. The complicated and poorly known wind and accretion structures of symbiotic binaries make simulating the low energy lines, e.g. He \textsc{i} and Balmer lines, a very difficult task, but the high ionisation energy of Fe$^{9+}$ means that the emission originates from very close to the WD, which means that this line is insensitive to the large scale wind structure.

\subsection{O VI 6830 \AA\ Raman feature}\label{sec:raman}

The broad O \textsc{vi} 6830 \AA\ feature is due to inelastic scattering of O \textsc{vi} 1032 \AA\ photons by hydrogen atoms. In this Raman scattering process the O $\textsc{vi}$ photons are absorbed by a hydrogen atom at the ground state ($1s^2$S), which is then excited to an intermediate state. A photon with $ \lambda \sim 6830$ \AA\ is then emitted and the hydrogen is left in an excited state ($2s^2$S) \citep{Schmid89}. This Raman scattering process requires an ionising source hot enough to produce O \textsc{vi} in the vicinity of large amount of neutral hydrogen. The Raman scattering cross-section is $7.5 \times \sigma _{\mathrm{T}} \approx 5 \times 10^{-24}$ cm$^{2}$, where $\sigma _{\mathrm{T}}$ is the Thomson cross-section \citep{Schmid89, Lee97}. For an optical depth of 1, column densities of $N_H \approx 1/\sigma \approx 2 \times 10^{23}$ cm$^2$ are needed. Such densities are typically reached only in the innermost parts of the wind, or in the photosphere of the donor star (see also Fig.~\ref{fig:columnDensity}). 

The Raman scattered feature is observed almost exclusively in symbiotic binaries ($\approx $ 55\% of them in the Milky Way; \citealt{Akras19}), together with the 7088 \AA\ feature, which comes from Raman scattering of O $\textsc{vi}$ 1038 \AA\ (this is outside of the WiFeS R7000 wavelength range). Other Raman scattered lines, e.g. He $\textsc{ii}$ 1025 \AA\ $\rightarrow$ 6545 \AA\ \citep{Sekeras15}, have also been detected in some symbiotic systems, but these are mostly very faint and thus not detectable in our observations of \lin .


\section{\textsc{Cloudy} simulations setup}\label{sec:sim}

As described in Sec.~\ref{sec:data}, many of the observed lines in the high density nebulae around symbiotic binaries are products of complicated radiative and collisional processes, and thus explaining the source properties via these emission lines requires a detailed, simultaneous and self-consistent treatment of all the necessary complex processes. To tackle this problem, we used the open source spectral synthesis code \textsc{Cloudy}\footnote{www.nublado.org} version 17.01 \citep{Ferland17} to simulate the \lin\ system. 

We calculated a grid of photoionization models while assuming that the central ionizing source (the WD) emitted a blackbody spectrum, which is a reasonable approximation for the ionizing emission of nuclear burning WDs \citep{Woods16}. 
We have ignored the radiation from the giant star, because of its low effective temperature (4000 K). While the giant star dominates the infrared emission, its contribution to the optical line emission is very small compared to the white dwarf.

In our simulations the metallicity of the gas was set to one fifth of solar metallicity from \citet{Lee05, Grevesse10}, corresponding roughly to a typical SMC metallicity. 
A diffuse background radiation field was included in the calculations in the standard way it is implemented in \textsc{Cloudy}, where the radiation field shape and intensity were set to describe the cosmic radio to X-ray background \citep{Ostriker83, Ikeuchi86, Vedel94} with the cosmic microwave background included, and the extra heating and ionisation of cosmic rays were included in the calculations according to the mean ionisation rate of \citet{Indriolo07}.

\subsection{Density structure}

As mentioned in Sec.~\ref{sec:ov}, the ionising source in symbiotic binaries sits in the dense wind emitted by the giant donor star, which makes the density structure around the WD asymmetric. This makes simulating these complex objects with \textsc{Cloudy} problematic, because \textsc{Cloudy} is a one dimensional code. \lin\ and other symbiotic binaries have been previously studied with \textsc{Cloudy} \citep[see][for \lin]{Orio07}. However, our initial attempts to compute the nebula spectrum in 1D Cloudy calculations grossly failed in explaining our observations.  This is probably not surprising, given that the density profile towards the donor star is very different from the density profile in the  opposite direction.
For this reason, we constructed a model where we calculate the ionized gas structure with \textsc{Cloudy} along a number of paths from the WD and combine the results to get the 2D gas structure. We assume that the WD and the giant star are separated by 3.7 AU (see Sec.~\ref{sec:ov}) and that the mass-loss from the giant is spherically symmetric so that the ionization structure becomes rotationally symmetric along the axis between the WD and the giant star, and we can restrict the calculations to a 2D plane. 

The real situation is naturally more complex. The wind in AGB stars is driven by stellar pulsations and thus variable both in direction and time. However, the long-term average mass-loss is still often well approximated by a spherically symmetric formula \citep{Hoefner18}. Furthermore, hydrodynamical simulations \citep[e.g.][]{Mohamed07, Mohamed12,deValBorro09} show that the wind in symbiotic binaries is focused towards the binary orbital plane. Accurately accounting for this effect would require full 3D radiative transfer simulations, which is outside the scope of this paper. We note, however, that the wind focusing in this system appears to be moderate, given that our 2D simulations were able to correctly reproduce the global ionisation structure of the wind, as confirmed by the good consistency of the simulated spectra with observations, despite the richness and high statistical quality of the observational data. On the other hand, our initial experiments showed that these effects are escaping the 1D ionisation simulations which grossly fail in explaining the observed spectrum of LIN 358. Our calculations presented here are a significant improvement on the earlier 1D calculations and we have achieved consistent results at a significantly lower cost compared to what would be required for 3D simulations.

\begin{figure}
\centering
\includegraphics[width=0.47\textwidth]{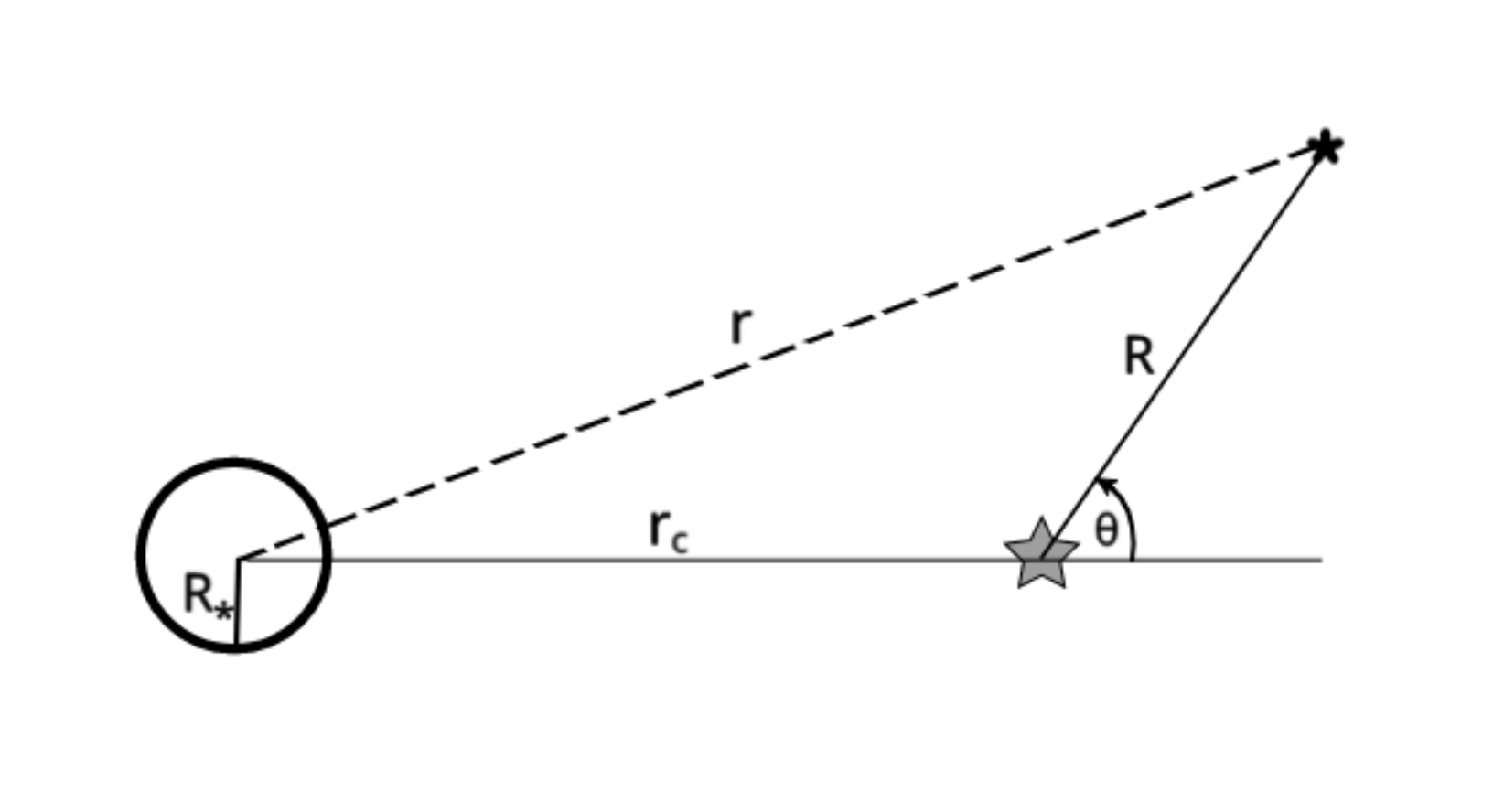}
\caption{The geometrical configuration of our simulations. The giant star and the centre of the spherically symmetric density distribution is marked with the black open circle. The white dwarf is marked with the grey star at the distance $r_c$ from the density centre. A test particle at distance $r$ from the centre of the density distribution will have coordinates $(R, \theta)$ in the WD-centred reference system.}
\label{fig:system}
\end{figure}

\begin{figure}
\centering
\includegraphics[width=0.47\textwidth]{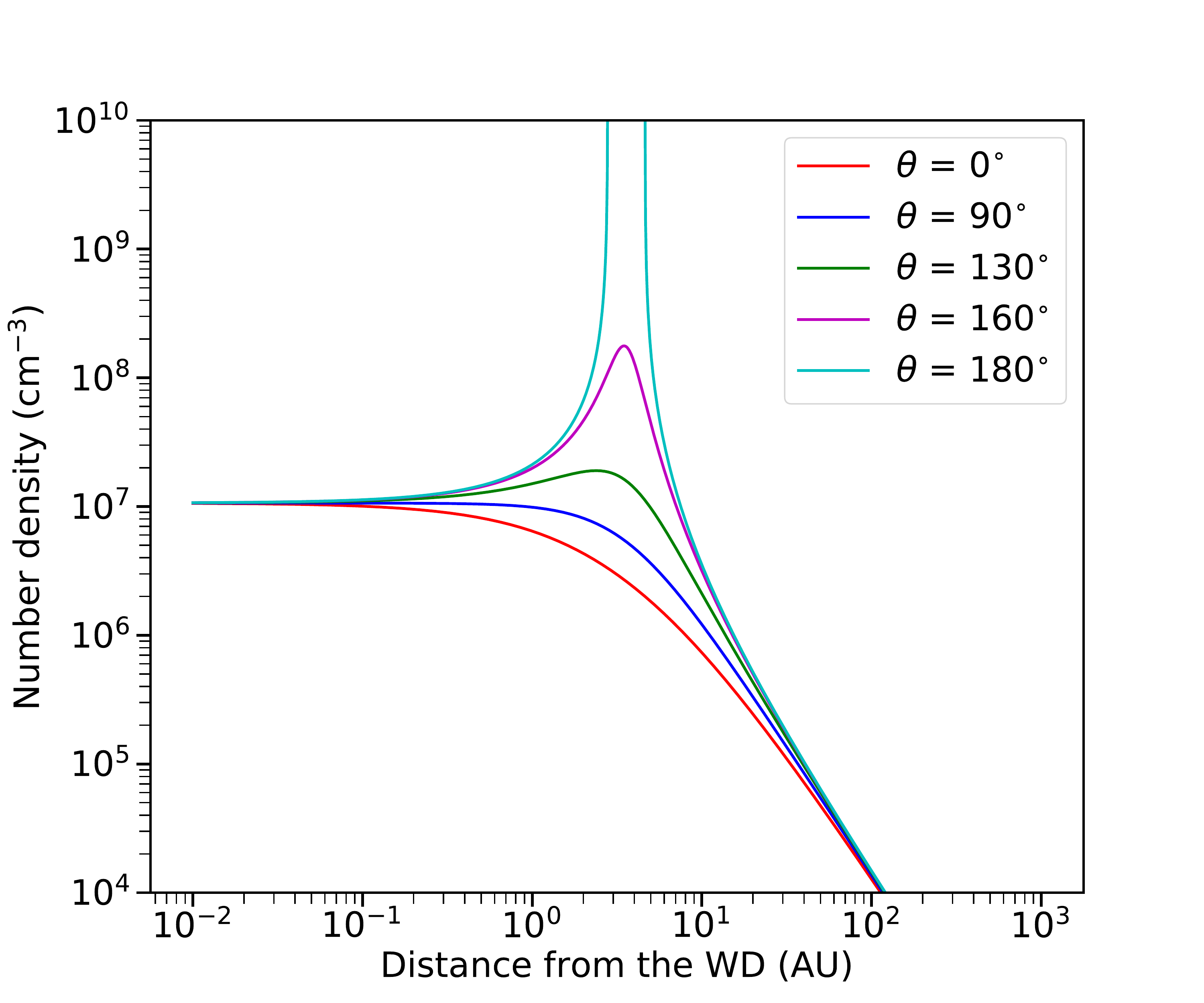}
\caption{The density distribution as seen by the WD with $n_c = 10^7$ cm$^{-3}$ and $r_c$ = 3.7 AU calculated for various angles. }
\label{fig:denstheta}
\end{figure}

We assume the mass-loss to be of the form
\begin{equation}\label{eq:massloss}
    \dot{M} = 4 \, \pi \, \mu \, m_{\mathrm{H}} \, v \, r^2 \, n(r) \, \left( 1 - \frac{R_*}{r} \right) ,
\end{equation}{}
where $\mu$ is the mean molecular weight, $m_{\mathrm{H}}$ is the mass of a hydrogen atom, $v = 15$ km s$^{-1}$ is the assumed constant wind velocity \citep{Chen17}, $R_{*} = 178$ R$_{\odot}$ is the origin of the stellar wind, i.e. the radius of the giant star, $r$ is the distance from the centre of the giant star, and $n(r)$ is the number density at distance $r$. The main consequence of Eq.~(\ref{eq:massloss}) is that the wind density structure follows a power-law ($r^{-2}$) distribution when $r \gg R_{*}$, but at the surface of the giant star the density becomes very high. We set the maximum density to $n = 10^{14}$ cm$^{-3}$ to avoid the infinite density at $r = R_{*}$. 
With this wind structure the ionising source is sitting off-centre at a distance $r_c$, but the density distribution from the white dwarfs point of view can easily be calculated following \citet{Arthur07}. First, we can write the density distribution in a form:
\begin{equation}\label{eq:nr}
    n(r) = n_c \, \left( \frac{r}{r_c} \right)^{-2} \, \left( 1 - \frac{R_*}{r} \right)^{-1} ,
\end{equation}{}
where $n_c$ is the number density at the distance $r_c$, i.e. at the position of the WD. We can now change the coordinates and centre the reference system to the position of the WD so that we can write
\begin{equation}\label{eq:rR}
    r^2 = R^2 + r_c^2 - 2 R r_c \mathrm{cos}(\pi - \theta) \, ,
\end{equation}{}
where $R$ is the distance measured from the WD to the direction of $\theta$, where the angle $\theta$ is measured from the symmetry axis between the WD and the giant star, and the giant star is in the direction $\theta = \pi$; see Fig.~\ref{fig:system}. Combining equations (\ref{eq:nr}) and (\ref{eq:rR}), we can write the density distribution from the white dwarfs point of view as 
\begin{equation}\label{eq:nRT}
    n(R, \theta) =  \frac{n_c \, r_c^2}{R^2 + r_c^2 + 2 R r_c \mathrm{cos}\theta }  \left( 1 - \frac{R_*}{R^2 + r_c^2 + 2 R r_c \mathrm{cos}\theta} \right)^{-1}.
\end{equation}{}
With Eq.~(\ref{eq:nRT}) we can calculate the density distribution along any line of sight, which is illustrated in Fig.~\ref{fig:denstheta} with $n_c = 10^7$ cm$^{-3}$ and $r_c = 3.7$ AU.

\section{Results}\label{sec:results}

Using the setup explained in Sec.~\ref{sec:sim} we performed simulations to see if we can reproduce the observed characteristics of the \lin\ by varying the main parameters of the problem: temperature $T_h$ and luminosity $L$ of the WD and the mass-loss rate $\dot{M}_{\mathrm{loss}}$ of the donor star. Strictly speaking, the WD luminosity is not an independent parameter. It is determined by the mass accretion rate modulo the regimes of the nuclear burning on the WD surface, whereas for the given binary system parameters, the mass accretion rate depends on the mass loss rate of the donor star. On the other hand, line luminosities scale with the gas density which in turn is proportional to the mass loss rate. Due to the non-linear nature of the problem, we  varied $\dot{M}_{\mathrm{loss}}$, $L$ and $T_h$ in an ad hoc iterative procedure to reach the values that best described the observed properties of \lin.  In this procedure, we considered  only the high excitation lines of He \textsc{ii} 4686 and \fe\ for which the line luminosities are robustly predicted in our simulations. As, in the relevant parameter range, the line luminosities are monotonic functions of temperature, WD luminosity and the donor mass-loss rate, the so found solution would be unique. Other significantly detected lines were used for an a posteriori consistency check.

\subsection{Colour temperature of the white dwarf}

\begin{figure}
\centering
\includegraphics[width=0.47\textwidth]{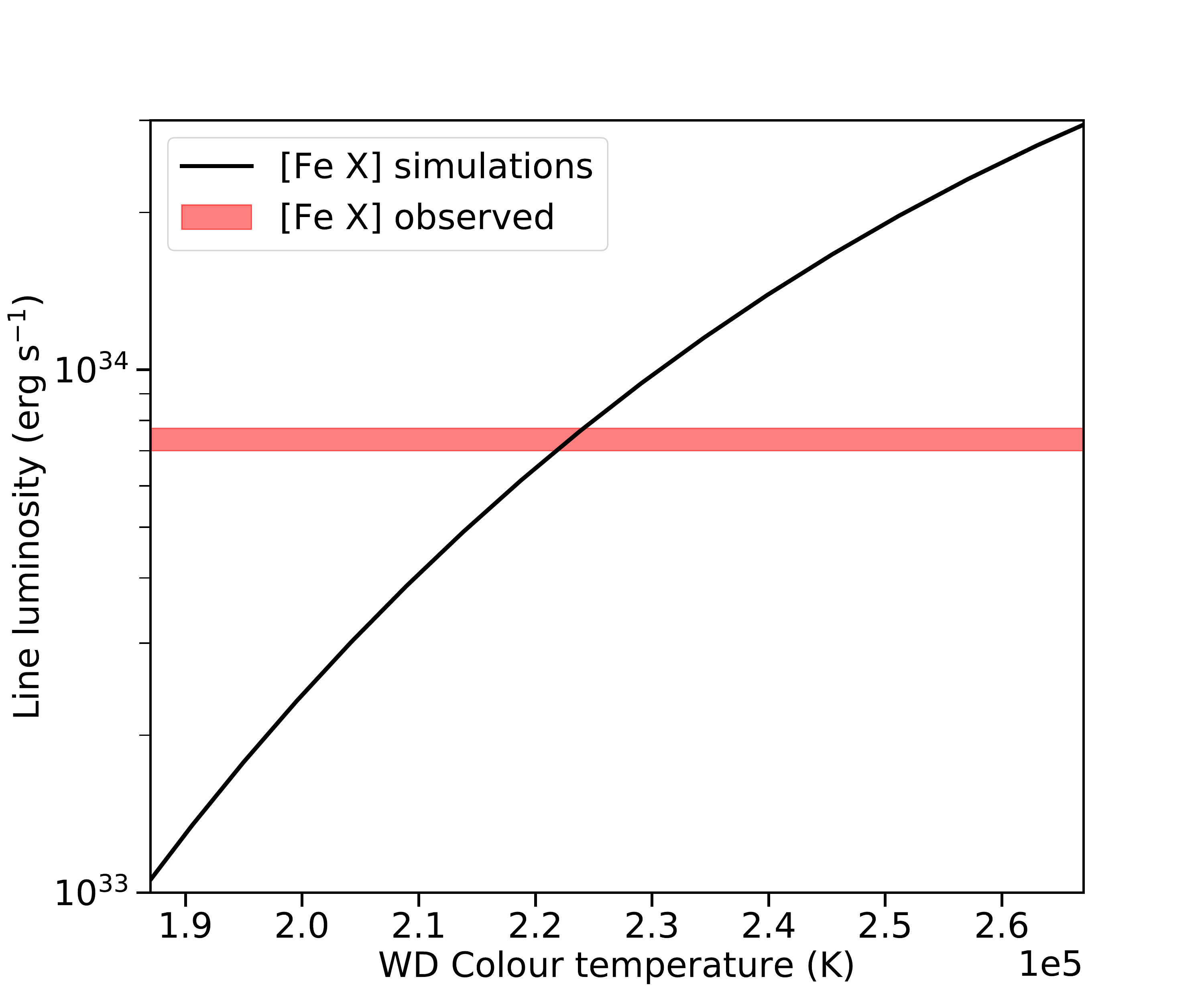}
\caption{Simulated \fe\ luminosity as a function of the WD blackbody temperature calculated with $L = 10^{38}$ erg s$^{-1}$ and $\dot{M}_{\mathrm{loss}} = 1.2 \times 10^{-6}$ M$_{\odot}$ yr$^{-1}$.}
\label{fig:FeXtemp}
\end{figure}

First, we determined the temperature of the ionising source by matching the observed \fe\ emission line to the simulations. As mentioned in Sec.~\ref{sec:fex}, this emission line is produced close to the white dwarf and  is  very sensitive to the temperature of ionizing radiation but insensitive to the structure  of the stellar wind at large distances from the white dwarf (see also Fig.~\ref{fig:massloss}).  The simulations and the comparison to the observed line luminosity is shown in Fig.~\ref{fig:FeXtemp}.  In this figure,  the WD luminosity and the wind mass loss rate are fixed at the values, determined in the above mentioned iterative procedure.  From this plot we can derive for the temperature:
\begin{equation}\label{eq:fextemp}
    T_h \, = \, ( \, 2.22 \pm 0.03 \, ) \, \times \, 10^5 \,\, \mathrm{K}  \,\,\, \approx \, 19 \,\,  \mathrm{eV},
\end{equation}
where the error correspond to the statistical error of the \fe\  line luminosity.  The so obtained value of the WD colour temperature is close to the temperatures $T_h = (2.275 \pm 0.3) \times 10^5$ K and $T_h = (2.50 \pm 0.1) \times 10^5$ K derived previously by \citet{Kahabka06} and \citet{Skopal15a}, respectively. In addition, this temperature is in agreement with the simple formula of \citet{Iijima81} used to estimate the effective temperature of a central source from the nebular emission line fluxes: 
\begin{equation}
    T (10^4 K) = 19.38 \sqrt{\frac{2.22 F_{4686}}{4.16 F_{H\beta} + 9.94 F_{4471}}} + 5.13,
\end{equation}
where $F_{4686}$, $F_{H\beta}$, and $F_{4471}$ are the fluxes of He $\textsc{ii}$ 4686, H $\beta$, and He $\textsc{i}$ 4471 emission lines, respectively. Using this formula for \lin, we get $T_{\mathrm{eff}} = 2.36 \times 10^5$ K. 
In all of the following calculations we have used the temperature derived from the \fe\ emission line in Eq.~(\ref{eq:fextemp}). 

\begin{figure}
\centering
\includegraphics[width=0.47\textwidth]{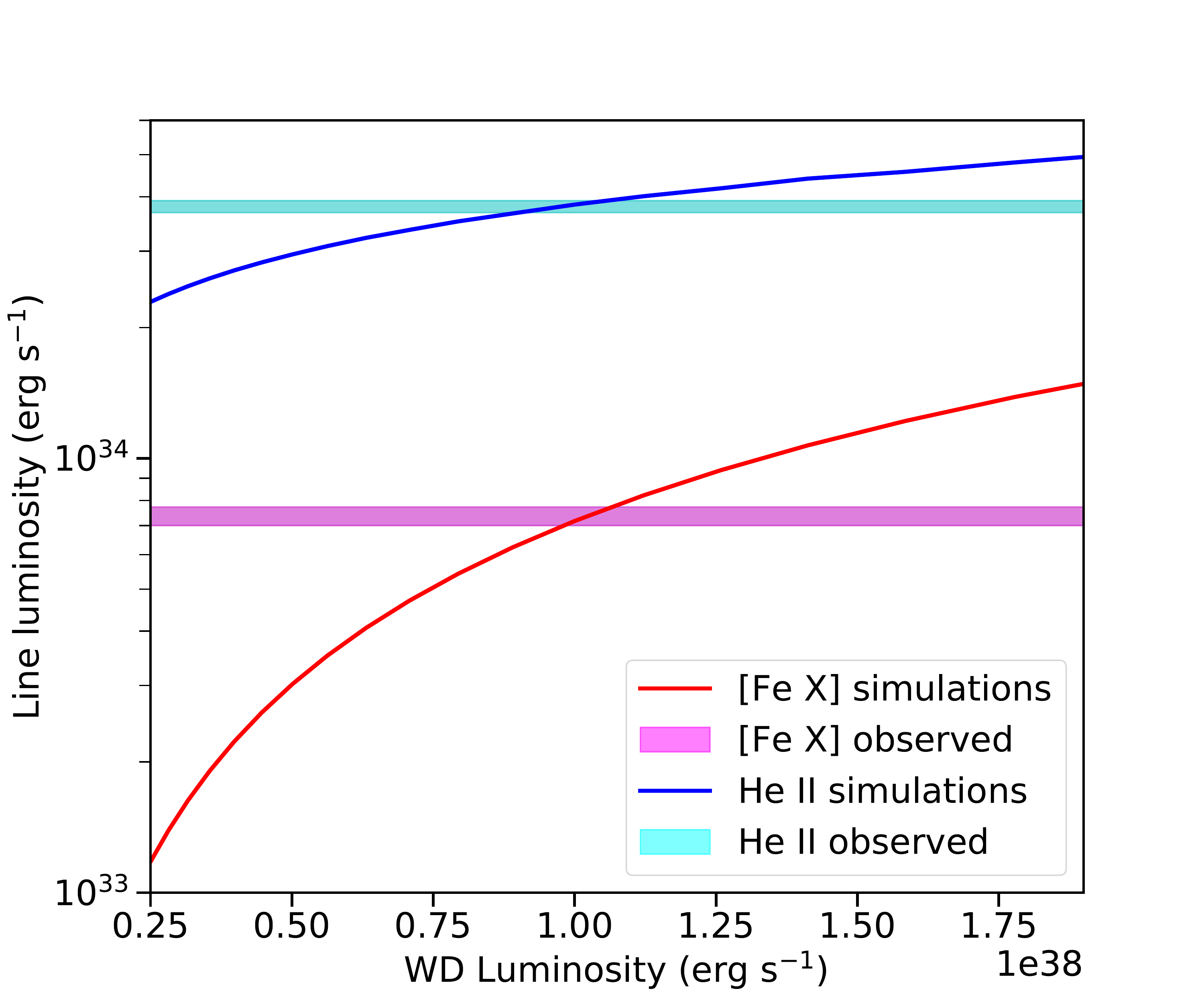}
\caption{Simulated \fe\ (red) and He \textsc{ii}  4686 \AA\ (blue) line luminosity  as a function of the WD luminosity for $T = 19$ eV and $\dot{M}_{\mathrm{loss}} = 1.2 \times 10^{-6}$ M$_{\odot}$ yr$^{-1}$.}
\label{fig:luminosity}
\end{figure}

\subsection{White dwarf luminosity}

To justify the choice of the WD luminosity we used the two high-excitation lines \fe\ and He \textsc{ii} 4686 \AA. The comparison between simulations and observations is shown in Fig.~\ref{fig:luminosity}, demonstrating that the two lines give consistent estimates of the WD luminosity. From this we can derive the WD luminosity:
\begin{equation}
    L \, = \, ( \, 1.02 \, \pm \, 0.15 \,) \, \times \, 10^{38} \,\, \mathrm{erg} \,\, \mathrm{s}^{-1}.
\end{equation}
This value is consistent with the previous values of \citet{Kahabka06} and \citet{Skopal15a}.

\subsection{Mass-loss rate of the donor star}

\begin{figure*}
\centering
\includegraphics[width=\textwidth]{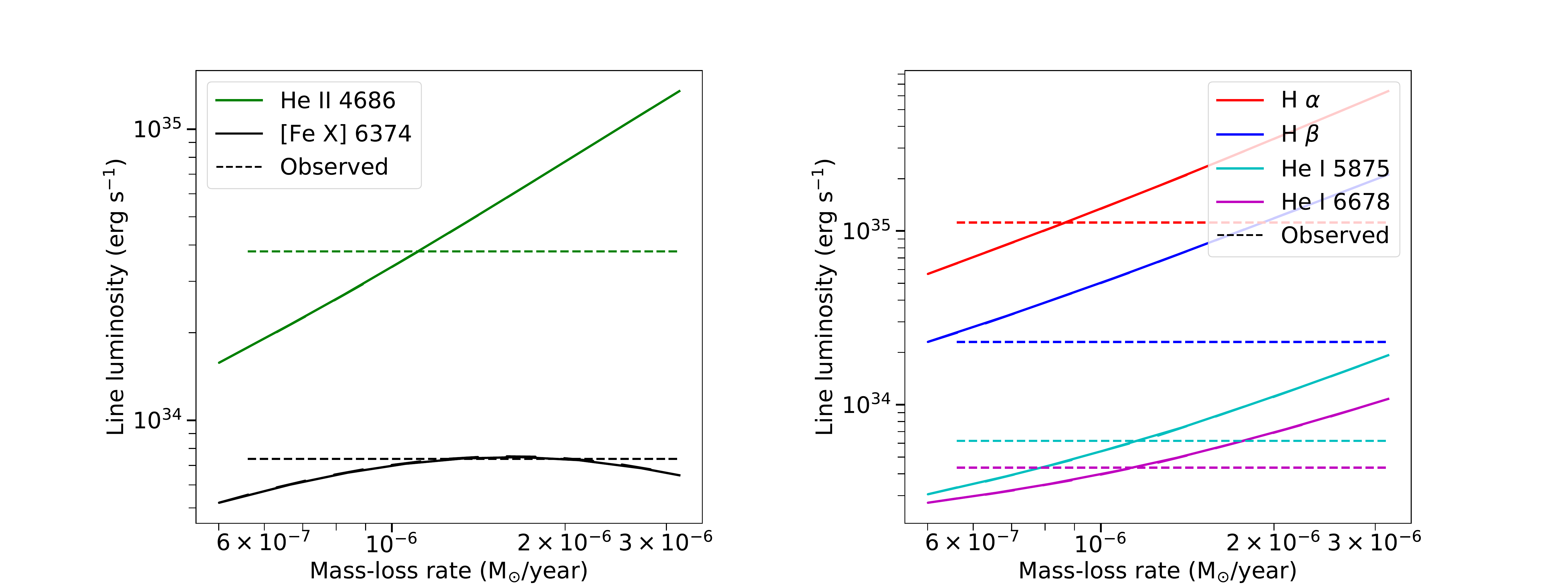}
\caption{The observed and simulated line luminosities for main emission lines in the spectrum of \lin\ shown as a function of the mass-loss rate of the giant star.  In the left panel we show the high excitation lines of He \textsc{ii} 4686 and \fe\ which luminosities are robustly predicted in our simulations. In the right panel we show the \ha , \hb , He \textsc{i} 5875, and He \textsc{i} 6678 \AA\ lines which are more sensitive to the collisional effects, self-absorption and detail of the wind structure near the donor star. The solid lines are predicted by the \textsc{Cloudy} simulations and the dashed horizontal lines show the observed values from Table~\ref{table:lums}. The mass-loss rate on the x-axis is from the Eq.~(\ref{eq:massloss}) with $v = 15$ km s$^{-1}$, $r=3.7$ AU, and $R_* = 178$ R$_{\odot}$.  }
\label{fig:massloss}
\end{figure*}

The density structure of the wind is related to the mass-loss rate from the donor star via Eq.~(\ref{eq:massloss}) with $v = 15$ km s$^{-1}$, $r=3.7$ AU, and $R_* = 178$ R$_{\odot}$. Simulated and observed line luminosities of the two high excitation lines of He \textsc{ii} 4686 and \fe\ are shown in the left panel of Fig.~\ref{fig:massloss}.
From this figure one can see that the luminosities of our two main diagnostic lines can be explained with a consistent mass-loss rate of 
\begin{equation*}
    \dot{M}_{\mathrm{loss}} \approx 1.2 \times 10^{-6} \,\, \mathrm{M_{\odot} \, yr^{-1}}
\end{equation*}

The right hand panel in Fig.~\ref{fig:massloss} shows  luminosities of  other principal emission lines. As one can see from this plot, the He \textsc{i} lines are consistent with the high excitation lines, but the Balmer lines show some scatter in mass-loss rate values, which however is not dramatic and its origin is reasonably well understood, as discussed in Sec. \ref{sec:discussion}.

\subsection{Structure of the emission regions of principal lines}

\begin{figure*}
  \centering
  \vbox{
  \includegraphics[width=0.8\textwidth]{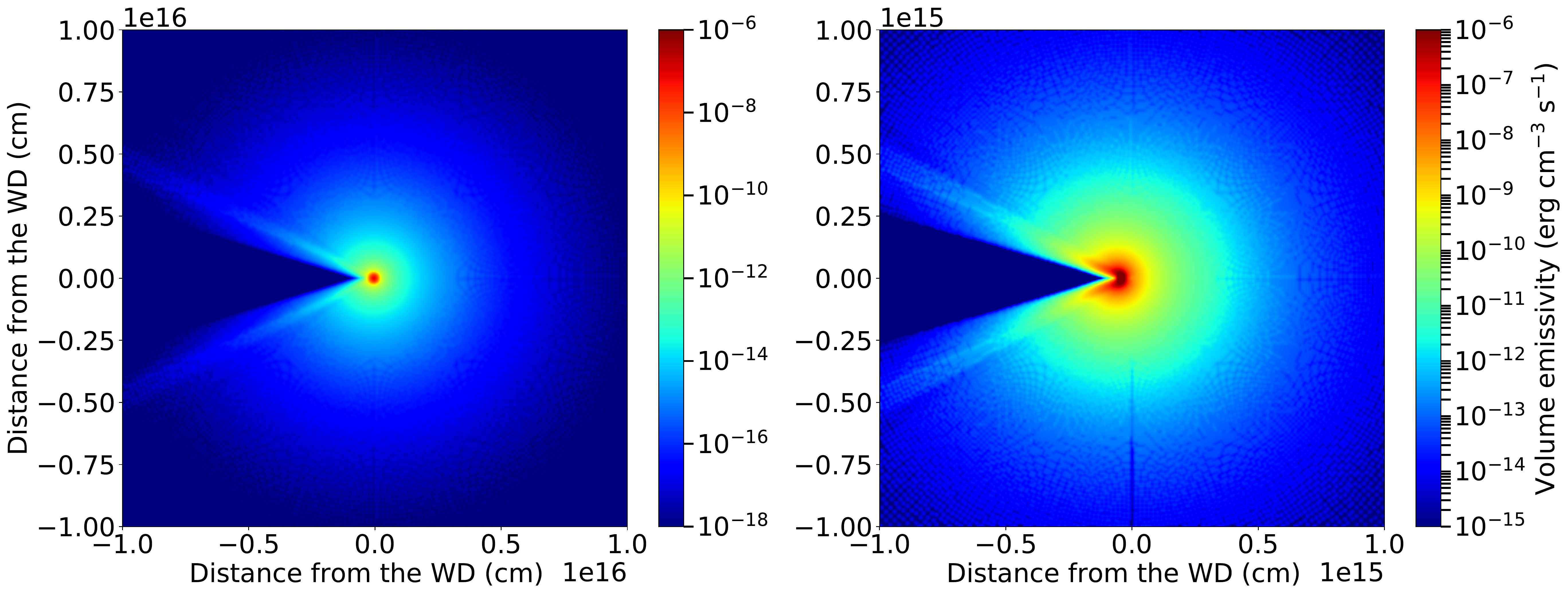}
  \vspace{0.5cm}
   \includegraphics[width=0.8\textwidth]{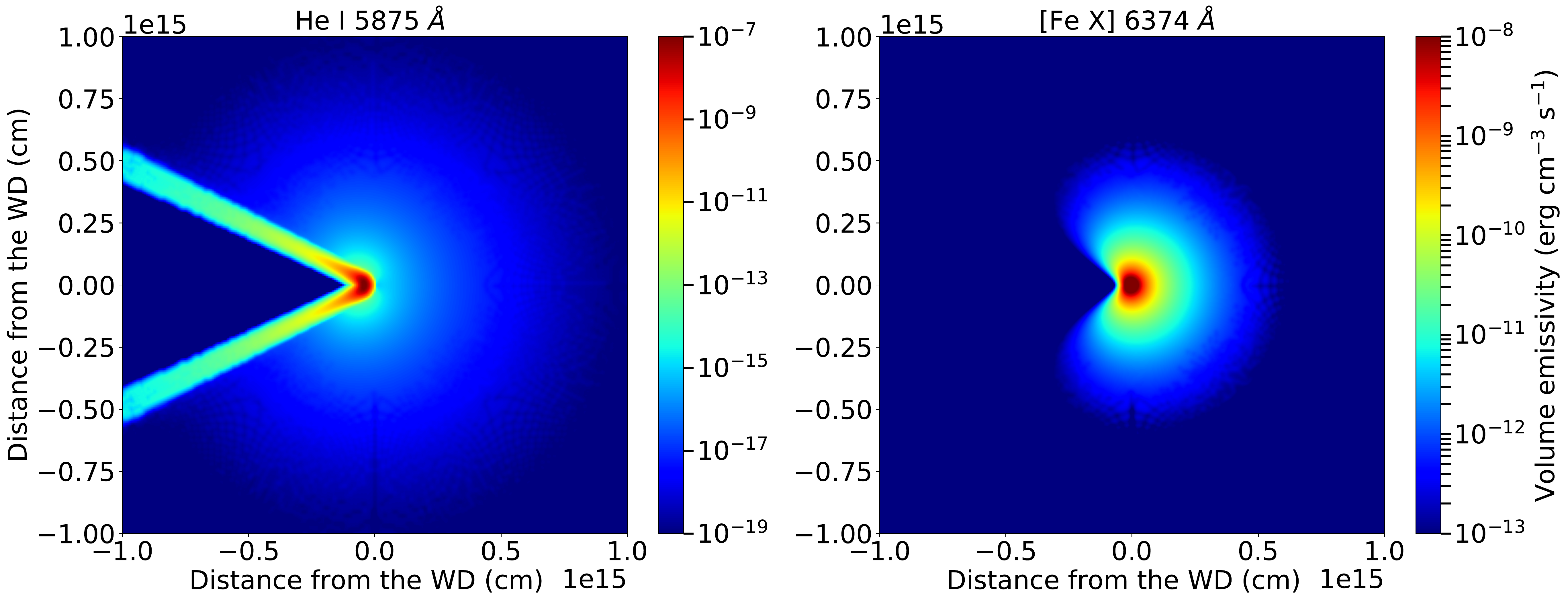}
   }
  \caption{Structure of the emission region in our best-fitting simulation (L = $10^{38}$ erg s$^{-1}$ and $\dot{M}_{\mathrm{loss}} = 1.2 \times 10^{-6}$ M$_{\odot}$ yr$^{-1}$) for different emission lines. The top row shows emissivity of H$\alpha$ line with two different length scales of $10^{16}$ cm (left)  and $10^{15}$ cm (right). The lower panels show emissivity of He \textsc{i} on the left and of \fe\ on the right at the length scale of $10^{15}$ cm. The colours represent the line volume emissivity in erg s$^{-1}$ cm$^{-3}$ according to the colour-bars. In this representation the WD is located at the centre of the image at coordinates (0, 0) and the donor star is to the left of the WD at a distance of 3.7 AU (5.5 $\times$ $10^{13}$ cm). }
  \label{fig:emissivities}
\end{figure*}

With our simulations we can investigate where the majority of the nebular emission in a symbiotic binary originates. This is illustrated in the top two panels in Fig.~\ref{fig:emissivities}, where we show the volume emissivity of several emission lines. The emissivity is shown in an arbitrary plane, because we have assumed an azimuthal symmetry. From these figures one can see that the emissivity distribution for low excitation lines  is quite  asymmetric along the line connecting the white dwarf and the donor star. Indeed, in the simple Str\"omgren sphere case, the radius of the ionized region in a constant density nebula containing only hydrogen is:
\begin{equation}\label{eq.stromgren}
\mathrm{R_S} = \left( \frac{3}{4 \pi} \frac{\mathrm{\dot{N}_{ph}}}{\mathrm{n^2} \alpha} \right) ^{\frac{1}{3}} 
 \approx 155 \mathrm{AU} \left( \frac{\mathrm{\dot{N}_{ph}}}{10^{48} \mathrm{s}^{-1}} \right) ^{\frac{1}{3}} \left( \frac{\mathrm{n}}{10^7 \mathrm{cm} ^{-3}} \right) ^{-\frac{2}{3}},
\end{equation}
where $\alpha$ is the recombination coefficient, $\mathrm{\dot{N}_{ph}}$ is the number of ionizing photons per second, and $\mathrm{n}$ is the number density of the ISM. The lower density away from the giant star (i.e. increasing x-coordinate in Fig.~\ref{fig:emissivities}) causes the ionized region to be more extended in this direction than towards the giant star. The dark cone to the left is the ``shadow'' of the giant star, which blocks the emission from the WD propagating to the left. 

\begin{figure}
\centering
\includegraphics[width=0.4\textwidth]{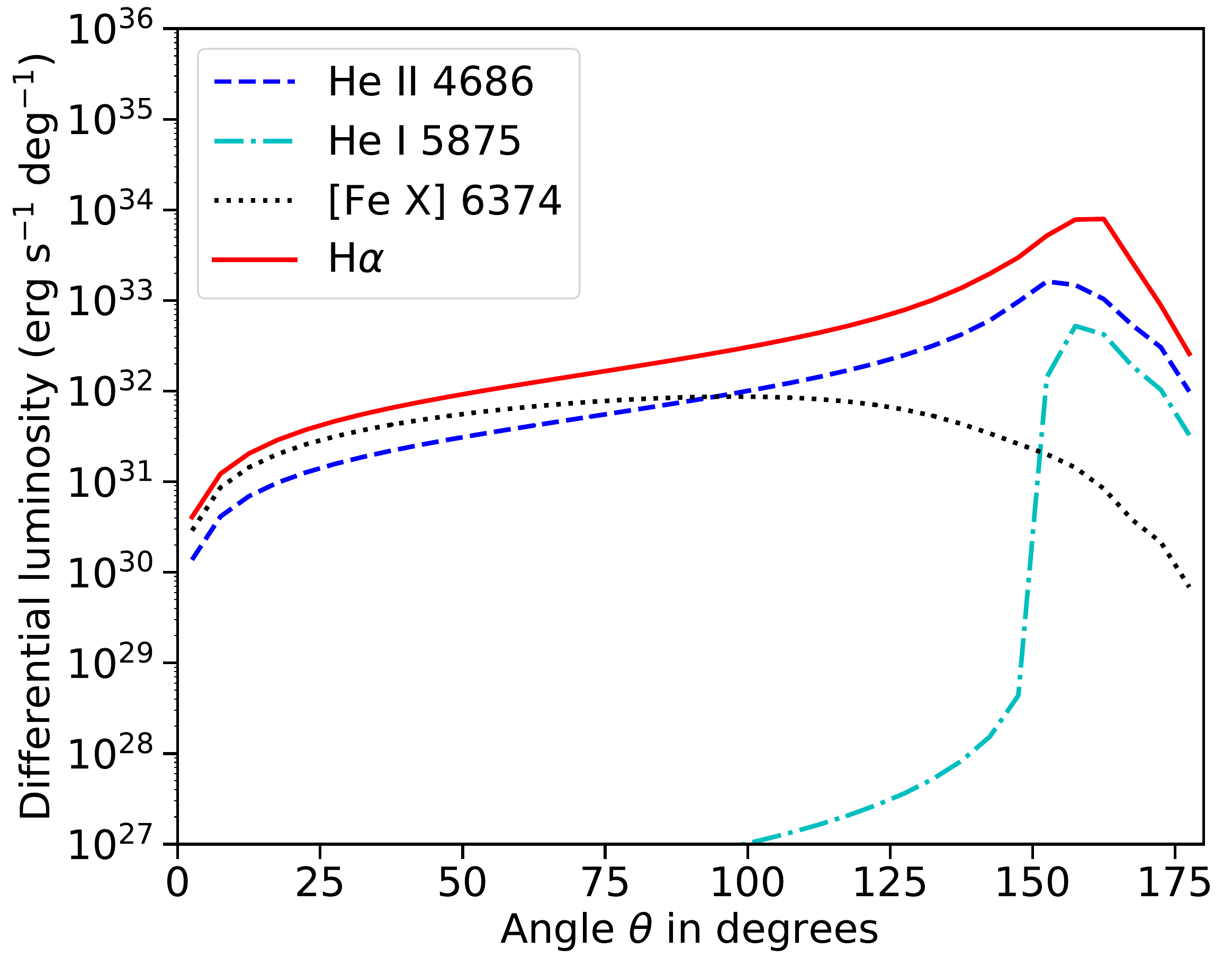}
\caption{Distribution of line luminosity around the white dwarf as a function of angle $\theta$ (see Fig.~\ref{fig:system}). The lines are calculated for L = $10^{38}$ erg s$^{-1}$ and $\dot{M}_{\mathrm{loss}} = 1.2 \times 10^{-6}$ M$_{\odot}$ yr$^{-1}$.}
\label{fig:diffLum}
\end{figure}

\begin{figure*}
\centering
\includegraphics[width=\textwidth]{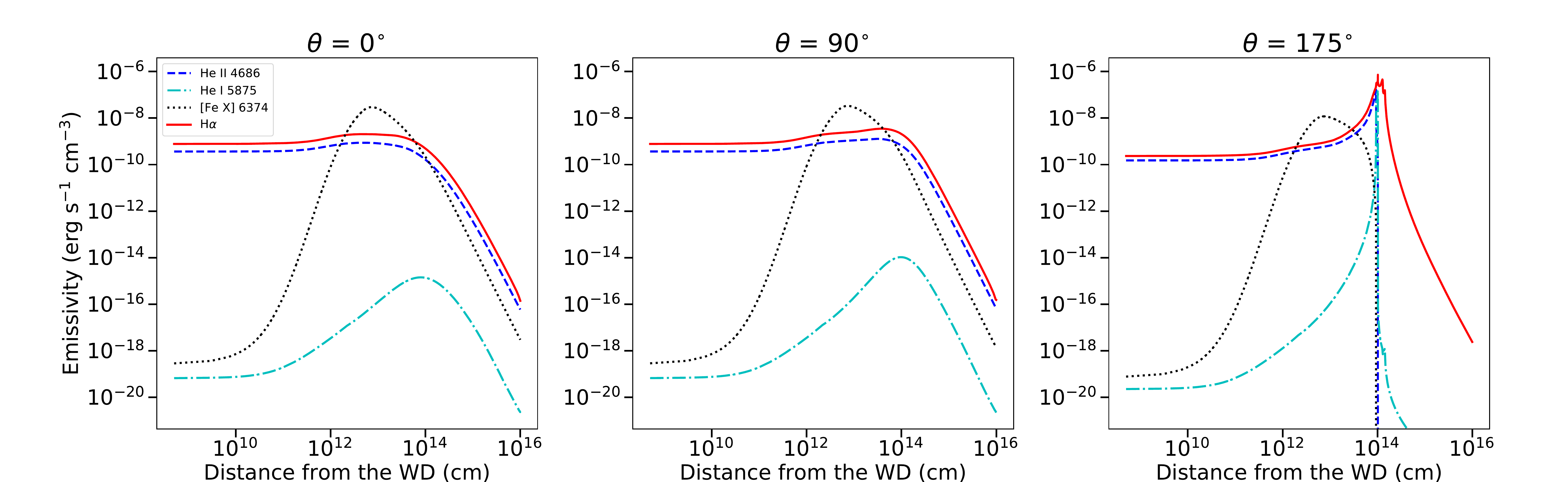}
\caption{The emissivity of various emission lines shown as a function of the distance from the ionising source with three different angles (see Fig.~\ref{fig:system}): $\theta = 0^{\circ}$ (left), $\theta = 90^{\circ}$ (middle), and $\theta = 175^{\circ}$ (right). The lines are calculated for L = $10^{38}$ erg s$^{-1}$ and $\dot{M}_{\mathrm{loss}} = 1.2 \times 10^{-6}$ M$_{\odot}$ yr$^{-1}$. }
\label{fig:em_radius}
\end{figure*}

From the Fig.~\ref{fig:emissivities} one can see that most of the emission originates from the vicinity of the giant star, where the density is higher. This is further illustrated by Fig.~\ref{fig:diffLum} where we show the simulated line luminosity as a function of the angle $\theta$ (see Fig.~\ref{fig:system}) for various emission lines. 
These figures also highlight the differences between different lines:  the \fe\ line originates roughly spherically from the immediate surroundings of the ionizing source, as already mentioned in Sec.~\ref{sec:fex}, whereas virtually all of the He \textsc{i} emission comes from a small region near the giant star, where the gas is not too highly ionized and the density is high enough that the collisional processes dominate the emission mechanisms. 
The emissivity as a function of the distance from the ionising source for several angles is shown in Fig.~\ref{fig:em_radius}. Also in this figure one can clearly see the same differences in the emission lines: the \fe\ emissivity peaks much to the WD than the H and He \textsc{I} lines, which all peak roughly at the same distance close to the giant star. For He \textsc{i} lines the emissivity peak is very sharp near the donor star, while for the others the emissivity is more extended.

\section{Discussion} 
\label{sec:discussion}

We have used 2D radiative transfer calculations based on the  \textsc{Cloudy} photoionization code to study ionization of the wind from the AGB donor star by the emission from the nuclear burning white dwarf in the LIN 358 system. As discussed in Sec.~\ref{sec:sim}, the real geometry of the wind in symbiotic binaries is not spherically symmetric and strictly speaking requires 3D simulations. However, our 2D simulations do capture correctly the main geometrical aspects of the photoionization nebula around the white dwarf -- significant asymmetry of the density distribution towards the donor star, away from it and in the direction normal to the orbital plane of the system. With this, our simulations successfully explain all the main emission lines except for the Balmer lines, which are modified by self-absorption and have complicated line profiles. Our approach enabled us to significantly improve on the 1D calculations (which in our runs failed to explain the observed spectrum) while also keeping the computational effort manageable. Based on our calculations we provide a self-consistent description of the observed optical spectrum and derived the main parameters of the problem -- wind mass loss rate of the AGB donor star, luminosity of the white dwarf and the colour temperature of its emission. 

The orbital parameters of LIN 358 are not known whereas the orbital separation of the system is an important parameter in our calculations. To this end, we used estimates based on numerical modelling and an analytical description of the wind Roche Lobe overflow accretion regime, as described in Sec.~\ref{sec:orb}. As these estimates bear some uncertainty we investigated the dependence of our results on the assumed binary separation $a$ and found that they are relatively insensitive to $a$ in the relevant range of values of $a$. In particular, we have found that increasing $a$ by a factor of two increases the derived mass-accretion rate by only $\sim$20\%, and the temperature and luminosity change by $\sim$2\%.  Similarly, decreasing the binary separation to 2 AU on the other hand decreases the mass accretion rate by $\sim$15\%. 

\subsection{Low excitation lines}

Although absolute luminosities of \ha\ and \hb\ lines can be roughly accounted in our simulations (Fig. \ref{fig:massloss}) their observed line ratio is not reproduced. The line ratio in our simulations remains constant at $\approx 3$, as expected from the Case B recombination \citep{Osterbrock06}, but the observed line ratio is higher, \ha /\hb\ $\approx$ 4.9. This higher Balmer decrement could have been caused by ISM dust absorption, but this is unlikely, given the small value of interstellar reddening towards the source (see Sec.~\ref{sec:obs}). The most likely reason for the high Balmer decrement is self-absorption, which is not fully captured by our simple simulation setup. 
High \ha /\hb\ line ratios have been previously observed in e.g. AGN and other symbiotic binaries, and our observed value can be explained by e.g. the calculations done by \citet{Netzer75}, but the complicated high density structure near the surface of the giant star and inside the wind acceleration radius, where the collisional effects play a major role in the production of Balmer lines, cannot be fully described by our simple simulation setup. The wind structure is likely more complicated than a smooth power-law profile and there can be clumps, inhomogeneities, and shock waves. In addition, the wind can be gravitationally focused towards the orbital plane of the binary \citep{deValBorro09, Shagatova16}, and simulating this would require full 3D calculations.

\begin{figure}
\centering
\includegraphics[width=0.47\textwidth]{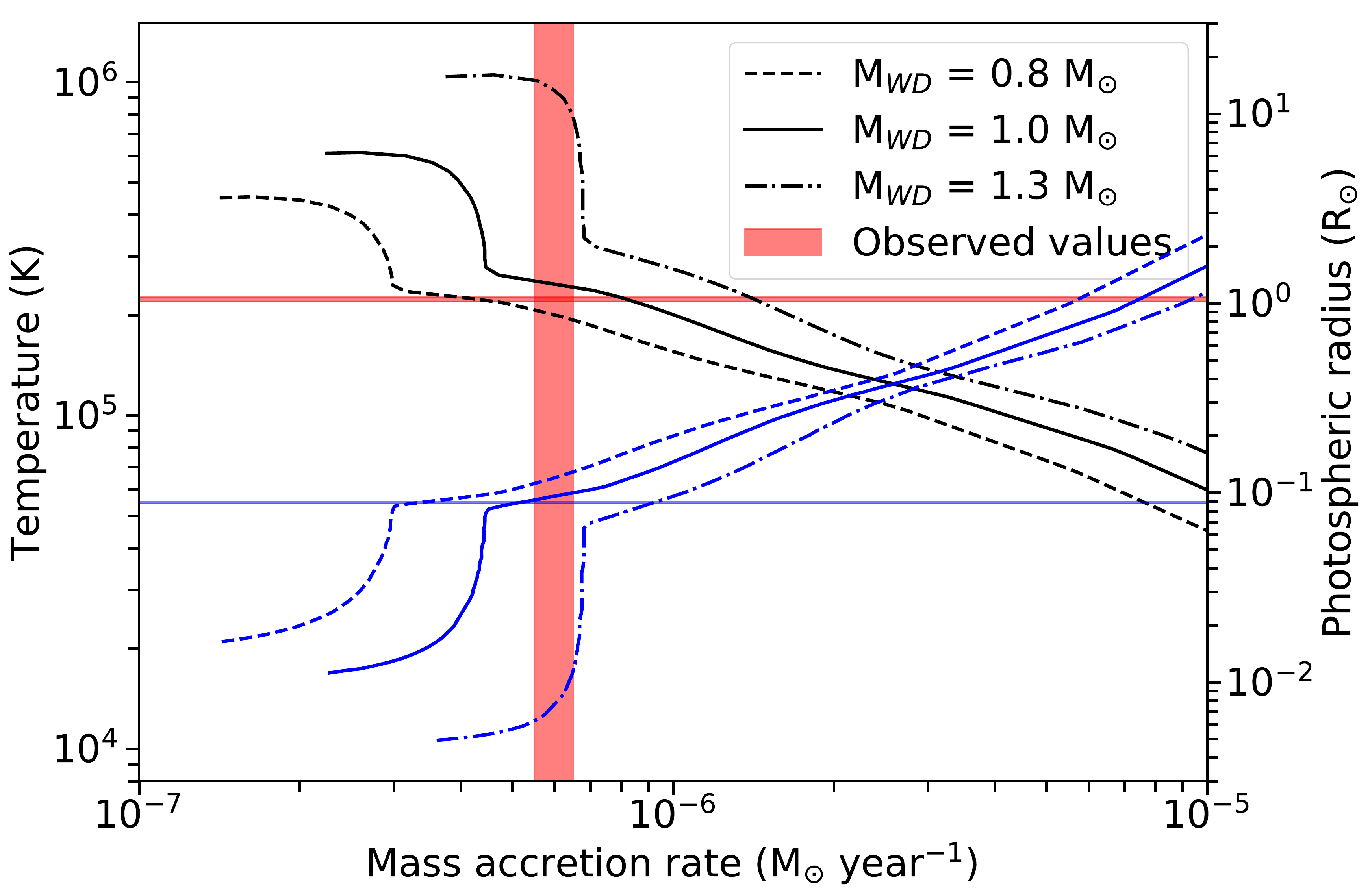}
\caption{The photospheric temperature (black lines) and radius (blue lines of the white dwarf shown as a function of the mass-accretion rate. The dashed, solid, and dash-dotted lines show the results of \citet{Hachisu99b} for WD masses of 0.8, 1.0, and 1.3 M$_{\odot}$, respectively. The red areas show our results for \lin , and the blue horizontal line shows the photospheric WD radius derived by \citet{Skopal15a}. }
\label{fig:massloss_temp}
\end{figure}

\subsection{Mass accretion rate}

Steady nuclear burning on the white dwarf surface can occur only in a rather narrow range of mass-accretion rates, below which the WD exhibits nova outbursts, and above which the WD is believed to have an expanded photosphere and lose the excess mass via high velocity winds \citep{Hachisu96, Nomoto07, Wolf13}.  

Our results suggest that the giant star in the \lin\ system is losing mass through stellar winds with a mass-loss rate of $\dot{M}_{\mathrm{loss}} \approx 1.2 \times 10^{-6}$ M$_{\odot}$ yr$^{-1}$. This value is well within the range of typical mass-loss rates for O-rich AGB stars like \lin\ \citep{Ramstedt09, Groenewegen18}.
How much of this material ends up on the WD is not fully clear, but using the same method as described by e.g. \citet{Abate18, Ilkiewicz20, Belloni20}
we can estimate the accretion efficiency to be $\approx 0.5$ which means the WD would accrete material at rate $\dot{M}_{\mathrm{acc}} \approx 6 \times 10^{-7}$ M$_{\odot}$ yr$^{-1}$. Recalling that for a 1 M$_{\odot}$ white dwarf stable nuclear fusion can occur in the range of mass accretion rates from $\approx 2 \times 10^{-7}$ M$_\odot$ yr$^{-1}$ to $\approx 4 \times 10^{-7}$ M$_\odot$ yr$^{-1}$ \citep{Nomoto07, Wolf13}, we conclude that the white dwarf in \lin\ accretes above the stability strip. In this regime, only a fraction of accreted material can be retained, the maximal rate  given  by the upper boundary of the stability strip, $\approx 4 \times 10^{-7}$ M$_\odot$ yr$^{-1}$ for a 1 M$_{\odot}$ white dwarf, the rest being blown away in a radiation driven wind \citep{Hachisu96, Nomoto07, Wolf13}. The high velocity WD wind will further complicate the geometry of the photoionization problem. 
Interestingly, the upper boundary of the stability strip, $\dot{M} = 4 \times 10^{-7}$ M$_{\odot}$ yr$^{-1}$, corresponds to the bolometric luminosity of  $L = 1.1 \times 10^{38}$ erg s$^{-1}$, which is very close to the luminosity of the white dwarf derived in our photoionisation calculations, $L = ( 1.02 \pm 0.15) \times 10^{38}$ erg s$^{-1}$. We emphasise that this value of WD bolometric luminosity was obtained in Sec.~\ref{sec:results} on completely different grounds. 

Conversely, we can use bolometric luminosity of the WD derived from photoionisation calculations to constrain the accretion efficiency in LIN358. Indeed, given the efficiency of the nuclear burning and assuming solar abundances, a luminosity of $1.02 \times 10^{38}$ erg s$^{-1}$ requires the mass accretion rate of $\dot{M} = 3.7 \times 10^{-7}$ M$_{\odot}$ yr$^{-1}$. Given the mass-loss rate of $\dot{M}_{\mathrm{loss}} \approx 1.2 \times 10^{-6}$ M$_{\odot}$ yr$^{-1}$ in \lin, the accretion efficiency is $ 0.31$. Taking into account that the bolometric luminosity is close to the upper boundary of the stability strip for 1 M$_{\odot}$ WD, this value should be considered as a lower limit. We also note that if \lin\ harbours a more massive white dwarf as suggested e.g. in Orio et al. (2007), the conclusion that the WD is accreting above the stable nuclear burning limit would not change as the value of $\dot{M} = 3.7 \times 10^{-7}$ M$_{\odot}$ yr$^{-1}$, corresponding to the luminosity of $1.02 \times 10^{38}$ erg s$^{-1}$, exceeds the lower boundary of the stability strip for any WD mass up to about 1.3 M$_{\odot}$. On the other hand, we argue that the mass of the WD in LIN358 is larger than 0.9 M$_{\odot}$ as for smaller masses the luminosity of $10^{38}$ erg s$^{-1}$ could not be maintained for extended period. 

Furthermore, we use the wind solution from \citet{Hachisu99b} to plot in Fig.~\ref{fig:massloss_temp} the photospheric temperature and radius of an accreting WD as a function of the mass-accretion rate. In this plot we also show our best fit WD colour temperature,  our estimate of the mass accretion rate and the WD photospheric radius measurement from \citet{Skopal15a}. As one can see, the lines cross  between the 0.8 and 1 $M_{\odot}$ curves derived from \citet{Hachisu99b} calculations.

Conversely, we can use bolometric luminosity of the WD derived from photoionisation calculations to constrain the accretion efficiency in LIN358. Indeed, given the efficiency of the nuclear burning and assuming solar abundances, a luminosity of $10^{38}$ erg s$^{-1}$ requires the mass accretion rate of $\dot{M} = 3.5 \times 10^{-7}$ M$_{\odot}$ yr$^{-1}$. Given the mass-loss rate of $\dot{M}_{\mathrm{loss}} \approx 1.2 \times 10^{-6}$ M$_{\odot}$ yr$^{-1}$ in \lin, the accretion efficiency is $\approx 0.3$. Taking into account that the bolometric luminosity is close to the upper boundary of the stability strip for 1 M$_{\odot}$ WD, this value should be considered as a lower limit. We also note that if \lin\ harbours a more massive white dwarf as suggested e.g. in Orio et al. (2007), the conclusion that the WD is accreting above the stable nuclear burning limit would not change as the value of $\dot{M} = 3.5 \times 10^{-7}$ M$_{\odot}$ yr$^{-1}$ corresponding to the luminosity of $10^{38}$ erg s$^{-1}$ exceeds the lower boundary of the stability strip for any WD mass up to about 1.3 M$_{\odot}$. On the other hand, we argue that the mass of the WD in LIN358 is larger than 0.9 M$_{\odot}$ as for smaller masses the luminosity of $10^{38}$ erg s$^{-1}$ could not be maintained for extended period.

\begin{figure}
\centering
\includegraphics[width=0.47\textwidth]{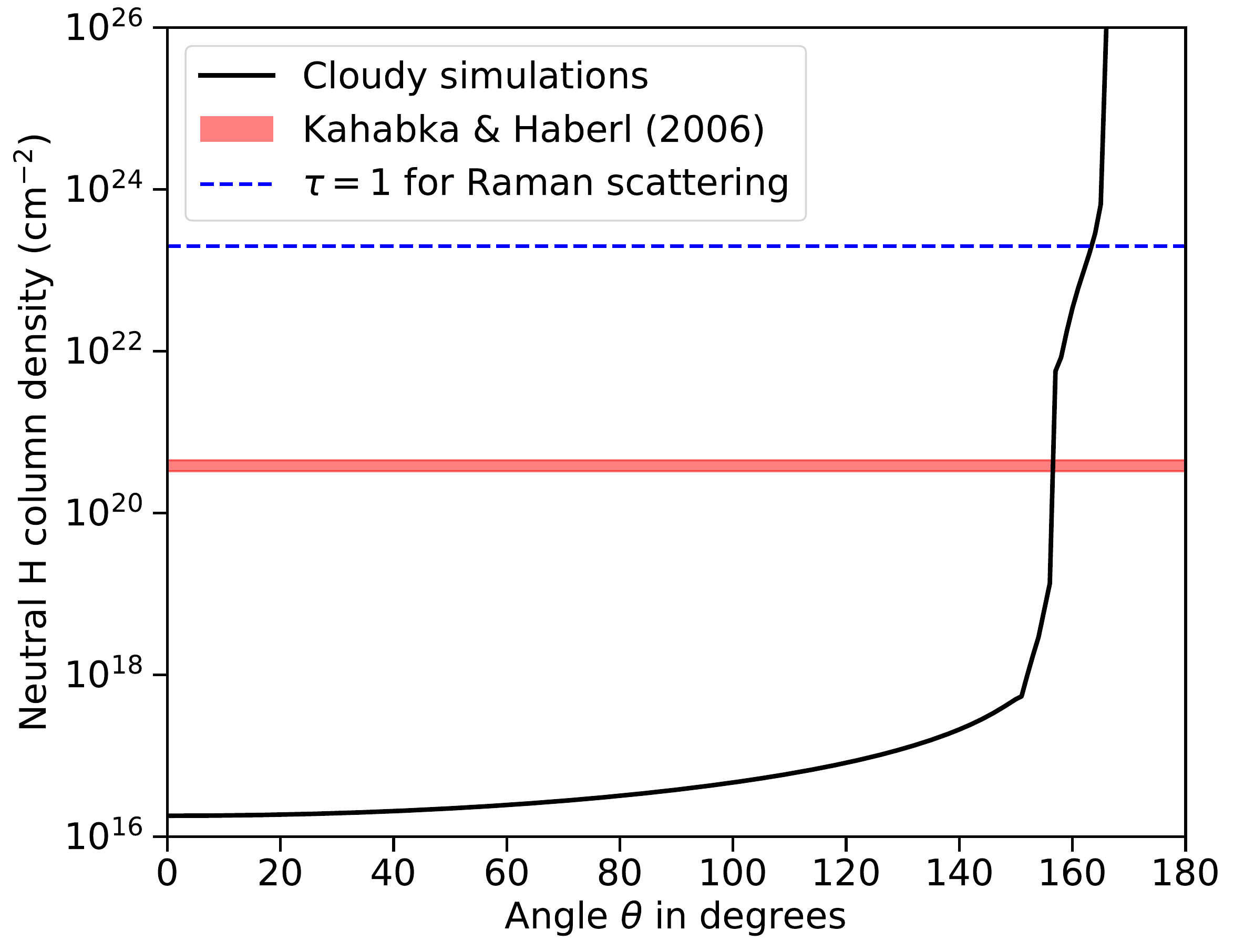}
\caption{The neutral hydrogen column density around the WD as a function of the angle $\theta$ (see Fig.~\ref{fig:system}) calculated for L = $10^{38}$ erg s$^{-1}$ and $\dot{M}_{\mathrm{loss}} = 1.2 \times 10^{-6}$ M$_{\odot}$ yr$^{-1}$. The black line shows the simulation results, red line shows the upper limit from \citet[][see text]{Kahabka06}, and the blue dashed line shows the column density at which the optical depth for Raman scattering equals unity (see Sec.~\ref{sec:raman}). }
\label{fig:columnDensity}
\end{figure}

Notably, any high velocity winds ($\sim$ 100's -- 1000 km s$^{-1}$) driven from the WD, as is predicted for accretion rates above the stable accretion regime \citep{Hachisu99b}, should prove strongly supersonic within the red giant wind medium. This would produce a strong shock, in a manner analogous to the colliding winds found in some high-mass binaries \citep[e.g.,][]{Dougherty2005}. Detecting such emission in \lin\ should be feasible; indeed, shock emission from colliding winds has previously been detected in a massive binary in the SMC \citep{Naze2007}. Modelling the additional emission expected from such a shock is, however, beyond the present scope; we address this, as well as prospects for constraining WD accretion physics, in a subsequent effort.

\subsection{Circumstellar absorption}
\label{sec:csmabs}

X-ray spectral fitting of LIN 358 gave the column density of neutral material intrinsic to SMC of $(3.9 \pm 0.6) \, \times 10^{20}$ cm$^{2}$ (on top of Galactic absorption of $3.7 \times 10^{20}$ cm$^{2}$) \citep{Kahabka06}. This number includes contributions from the neutral ISM in the SMC as well as the neutral circumstellar material (CSM) around \lin. In a more general context, attenuation by the dense wind from the giant star is often proposed to explain the paucity of observed symbiotic binaries. However, in many cases the wind should be too highly ionized to provide significant level of attenuation \citep{Nielsen15}. 

Our calculations show that this is also true for \lin, where the circumstellar material is mostly ionized except for a narrow cone towards the donor star. 
We show in Fig.~\ref{fig:columnDensity} the neutral hydrogen column density $\mathrm{N_H}$ obtained from our simulations as a function of the viewing angle from the WD. From this figure one can see that the wind is highly ionized everywhere except angles $\Theta \gtrsim 150^{\circ}$ where a large amount of neutral gas is present with column densities in excess of $\mathrm{N_H} \gtrsim 10^{22}$ cm$^{-2}$. However, the solid angle subtended by the regions of large column density is only about $\sim 0.07$ of $4\pi$. Therefore, even if the binary inclination angle is sufficiently small, large $\mathrm{N_H}$ values will only be observed in the narrow range of binary orbital phase, when the donor star is close to the line of sight.

We thus conclude that in the case of LIN 358, the circumstellar material can not provide any notable attenuation of the white dwarf emission, and that the excess absorption (above the Galactic value) observed in the X-ray spectrum of the source is due to the ISM in the SMC. Given the low colour temperature of LIN 358, even a modest amount of absorption by the neutral gas in the Milky Way and SMC is sufficient to attenuate its emission by a large factor.

Although in \lin\ the white dwarf emission freely escapes the surrounding CSM, we did not detect any extended ($\gtrsim 1$ pc) ionized nebula around the source. Super-soft X-ray sources are expected to ionize the surrounding ISM and create a distinct H \textsc{II} region around them \citep{Rappaport94, Remillard95, Woods16}. The expected presence of such ionized nebulae around accreting WDs has been used to constrain their accretion history and their role in producing  Type Ia supernovae \citep{Woods16, Woods17, Woods18, Graur19, Kuuttila19, Farias20}, but the ambient ISM density around \lin\ appears to be too low for a nebula to be detectable.

\subsection{LIN 358 in the context of the origin of SN Ia}

Our results suggest that the white dwarf in \lin\ is  growing in mass with a rate of $\approx 4 \times 10^{-7}$ M$_{\odot}$ yr$^{-1}$. For this reason \lin\ and symbiotic systems in general have been considered as prospective SN Ia progenitors \citep[see][for a review]{Maoz14}. One of the well known problems of this scenario, limiting the possible contribution of symbiotic systems to SN Ia production, is the short lifetime of an AGB star, of the order of $\sim 10^5$ years \citep{Yungelson98}.
Indeed, even a WD accreting above the steady burning limit and growing its mass at the maximum rate, similar to \lin, would gain only $\sim 0.04$ M$_{\odot}$ in the course of the symbiotic phase. The WD would have to be initially very massive in order to reach the Chandrasekhar mass, but such massive WDs are thought to be formed as ONeMg-rich WDs, which are thought to rather form a neutron star via AIC instead of exploding \citep{Nomoto84, Nomoto91}. The explosion and collapse mechanisms are however not yet fully understood and recent simulations show that ONeMg-rich WDs can explode \citep[see e.g.][]{Marquardt15, Jones16, Jones19}.
Nevertheless, symbiotic binaries are thought of as possible progenitor candidates for some peculiar SNe Ia, especially those exhibiting signatures of interaction with hydrogen-rich circumstellar material \citep[SNe Ia-CSM; e.g. SN 2002ic][]{Hamuy03}. These supernovae exhibit strong early time \ha\ emission, consistent with the supernova ejecta interacting with dense circumstellar material \citep{Silverman13a}. In addition, SNe Ia-CSM show large \ha / \hb\ ratios \citep[e.g. $> 7$ for PTF11kx;][]{Silverman13b} caused by collisional excitation and Balmer self-absorption in a high density gas, similar to \lin\ and many other symbiotic binaries.

While an AIC is a more likely outcome of a symbiotic binary than a SN Ia \citep{Yungelson10}, given that the highest mass WDs formed are ONeMg-rich, 
the formation of neutron stars via AIC from AGB+WD binaries faces the same issue as SNe Ia progenitors, which is the short lifetime of the AGB star. The number of AIC progenitors with an AGB donor in the Galaxy has recently been estimated to be $\sim 30$ by \citet{Ruiter19}, who assumed a BHL wind description and a time-averaged accretion rate of $\sim 10^{-8}$ M$_{\odot}$ yr$^{-1}$. However, as shown here, the accretion rate can be much higher than that, meaning that the AGB+WD binaries like \lin\ can form neutron stars more efficiently. 

As mentioned above, however, symbiotic binaries can be difficult to detect, especially in X-rays, while in other wavelengths they can be difficult to separate from other astrophysical sources. For example, in the infrared the symbiotic binaries are dominated by emission from the AGB star and thus they can be difficult if not impossible to separate from e.g. single AGB stars. The best wavelength range to unambiguously detect symbiotic binaries may be the optical spectrum, because symbiotic binaries like \lin\ are often very bright in the He \textsc{ii} 4686 \AA\ emission line, which is a clear signature of an accreting white dwarf. This line, together with some possible high excitation state forbidden lines such as \fe, can be used to identify possible symbiotic candidates, but perhaps the most important identifiers are the Raman scattered O \textsc{vi} features at 6830 \AA\ and 7088 \AA . These features are observed almost exclusively in symbiotic binaries, and in the Milky Way the presence of the Raman scattered lines is confirmed in about 55\% of the symbiotic population; in the SMC the percentage is 92\% \citep{Akras19}. Future surveys focusing on these Raman features, like the RAMSES II search \citep{Angeloni19}, can shed more light on the true population of symbiotic binaries and therefore provide some constraints on the birthrates of SNe Ia and AICs from the symbiotic channel.


\section{Conclusions}
We have examined the properties of the SMC symbiotic binary \lin\ by comparing our optical spectroscopic observations with 2D radiative transfer simulations performed with the help of the \textsc{Cloudy}  photoionization code. Comparing the results of our simulations and observations, we have determined the colour temperature of the WD to be $T =  (2.23 \pm 0.03) \times 10^5$ K, its bolometric luminosity of $L = (1.02 \pm 0.15) \times 10^{38}$ erg s$^{-1}$, and the  mass-loss rate of the donor star to be $\dot{M}_{\mathrm{loss}} = 1.2 \times 10^{-6}$ M$_{\odot}$ yr$^{-1}$. 
Assuming a solar H to He ratio in the wind material, a lower limit to the accreted mass fraction in LIN358 is 0.31
We also determined the accretion rate on to the white dwarf to be $\dot{M}_{\mathrm{acc}} \approx 6 \times 10^{-7}$ M$_{\odot}$ yr$^{-1}$. These results indicate that the WD in \lin\ is accreting material at a rate above the stability strip of hydrogen fusion and may be loosing a fraction of the accreted mass via a high velocity wind. At these high accretion rates the photosphere of the white dwarf expands to a fraction of the solar radius, thus explaining the low colour temperature of the white dwarf emission. We speculate that many symbiotic systems may be operating in this regime, which explains the paucity of detected systems. For the mass of $\approx 1$ M$_{\odot}$, the white dwarf in \lin\ is growing its mass at the maximum possible rate of $\approx 4 \times 10^{-7}$ M$_{\odot}$ yr$^{-1}$. It is however unlikely that the white dwarf in \lin\ will ever reach the Chandrasekhar limit due to the short lifetime of the AGB donor star. Our calculations show that the circumstellar  material in \lin\ is nearly completely ionized everywhere except for a narrow cone around the donor star with an opening angle of 30 degrees, and the radiation of the white dwarf freely escapes from the system. The low energy absorption detected in the X-ray spectrum of this system is due to neutral ISM in the Milky Way and in the SMC.


\section*{Acknowledgements}

IRS and AJR are supported by the Australian Research Council through grant numbers FT160100028 and FT170100243, respectively. MG acknowledges partial support by the RSF grant 14-22-00271. TEW acknowledges support from the NRC-Canada Plaskett fellowship.

\section*{Data Availability Statement}

The data underlying this article will be shared on reasonable request to the corresponding author.



\bibliographystyle{mnras}
\bibliography{viitteet}

\bsp	
\label{lastpage}
\end{document}